\documentclass[final]{agujournal2019}
\usepackage{url}
\usepackage{lineno}
\usepackage[inline]{trackchanges}
\usepackage{soul}
\usepackage{amssymb}
\usepackage{amsmath}

\draftfalse

\journalname{Journal of Advances in Modeling Earth Systems (JAMES)}

\begin{document}

\title{A two-dimensional model for eddy saturation and frictional control in the Southern Ocean}

\authors{J. R. Maddison\affil{1}, D. P. Marshall\affil{2}, J. Mak\affil{3,4}, and K. Maurer-Song\affil{2}}

\affiliation{1}{School of Mathematics and Maxwell Institute for Mathematical
Sciences, The University of Edinburgh}
\affiliation{2}{Department of Physics, University of Oxford}
\affiliation{3}{Department of Ocean Science, Hong Kong University of Science and Technology}
\affiliation{4}{National Oceanography Centre, Southampton}

\correspondingauthor{J. R. Maddison}{j.r.maddison@ed.ac.uk}

\begin{keypoints}
\item {Presents a simplified model exhibiting eddy saturation and frictional control at equilibrium}
\item {Suggests a decadal variability time scale for the Antarctic Circumpolar Current}
\item {Suggests eddy energy is relatively sensitive to stochastic wind forcing}

\end{keypoints}

\begin{abstract}
The reduced sensitivity of mean Southern Ocean zonal transport with respect to surface wind stress magnitude changes, known as eddy saturation, is studied in an idealised analytical model. The model is based on the assumption of a balance between surface wind stress forcing and bottom dissipation in the planetary geostrophic limit, coupled to the GEOMETRIC form of the Gent--McWilliams eddy parameterisation. The assumption of a linear stratification, together with an equation for the parameterised domain integrated total eddy energy, enables the formulation of a two component dynamical system, which reduces to the non-linear oscillator of Ambaum and Novak (Q. J. R. Meteorolog. Soc. 140(685), 2680--2684, 2014) in a Hamiltonian limit. The model suggests an intrinsic oscillatory time scale for the Southern Ocean, associated with a combination of mean shear erosion by eddies and eddy energy generation by the mean shear. For Southern Ocean parameters the model suggests that perturbing the system via stochastic wind forcing may lead to relatively large excursions in eddy energy.
\end{abstract}

\section*{Plain Language Summary}
The Southern Ocean volume transport is linked to the global stratification to the north of the Southern Ocean. It is thus of interest to understand how the Southern Ocean volume transport responds to changes in the forcing. Eddy saturation in this case refers to the weak sensitivity of the Southern Ocean volume transport to changes in wind forcing, and this phenomenon is investigated in this work within the context of an idealised but analytically and mathematically tractable model. The model mathematically formalises the physical arguments presented in previous works, and leads to closed form expressions for model response time scales, namely an oscillation and decay time scale. Random but sustained perturbations can be included in the model, and statistics of the model response can still be derived analytically. The simple model here advances our understanding of the governing processes related to the phenomenon of eddy saturation, with possible implications for understanding the ocean's global overturning circulation.


\section{Introduction}

Numerical models with an explicit representation of mesoscale eddies indicate that the Southern Ocean may be eddy saturated, with a mean zonal transport which is relatively insensitive to changes in the magnitude of the mean surface wind stress \cite<e.g.,>[]{munday2013,farneti2015,bishop2016,hewitt2020}. However, in models with parameterised eddies that make use of classic forms of the Gent--McWilliams parameterisation \cite{gent1990,gent1995}, eddy saturation is often not observed. The Gent--McWilliams scheme captures the response of baroclinic eddies to a baroclinically unstable lateral density gradient, in the form of an eddy induced overturning that opposes the density gradients which are the source of the instability. However the eddy response is constrained by properties of the eddy field, and in particular by the eddy energy.

Inclusion of eddy energy information in mesoscale eddy parameterisations has become increasingly widespread, finding use in constraining the parameterised momentum forcing \cite<e.g.,>[]{jansen2014,bachman2019,bagaeva2024,yankovsky2024} and in variants of the Gent--McWilliams scheme \cite<e.g.,>[]{cessi2008,eden2008,marshall2010,mak2018,jansen2019}. The principal focus of the present article is on the GEOMETRIC scheme \cite{marshall2012,mak2018}, which makes use of an energetically constrained form of the Gent--McWilliams coefficient, with the constraint derived (and holding exactly) in the quasi-geostrophic limit \cite{marshall2012}. The GEOMETRIC scheme is closed via the definition of a parameterised eddy energy equation. Numerical applications of the GEOMETRIC scheme in various forms have resulted in eddy saturation in a zonally averaged planetary geostrophic channel model \cite{mak2017}, as well as three-dimensional primitive equation models in both idealised \cite{mak2018} and realistic global configuration settings \cite{mak2022}.

A simple two-layer analytical model for the Southern Ocean is formulated in \citeA{straub1993}, making an assumption of weak flow at depth. An increase in mean surface wind stress forcing leads to an increased Ekman transport in the upper layer, which must be counteracted by the influence of the eddies. In order for these to exist it is necessary for the flow to be baroclinically unstable, and the condition of critical flow stability implies a mean zonal transport that is independent of surface wind stress forcing magnitude.

A physical description of eddy saturation, which considers the additional element of eddy energetics, is provided in \citeA{marshall2017}. The description assumes a balance between surface wind stress and bottom form stress, connected by interfacial eddy form stress. The magnitude of the eddy form stress is controlled by the eddy energy, using the GEOMETRIC energetic bound. The argument is closed with a simple parameterised integrated eddy energy budget, with eddy energy generation balanced by a linear damping of eddy energy. Combined, this description predicts that it is the eddy energy, and not the mean zonal thermal wind transport, which increases when the mean surface wind stress is increased. This description moreover predicts ``frictional control'', as an increase in eddy energy dissipation must be balanced by an increase in eddy energy generation, which is achieved by an increase in mean vertical velocity shear and hence mean zonal thermal wind transport. The description in \citeA{marshall2017} is related to that in \citeA{straub1993}, except that a mean zonal thermal wind transport is selected based upon consideration of eddy energetics, rather than critical flow stability.

In \citeA{ambaum2014} a two-dimensional dynamical system for atmospheric storm tracks is described, based upon the principles of an erosion of baroclinicity due to eddy heat fluxes, and a growth in eddy heat flux due to baroclinic instability (or the reverse of each). These are precisely the physical mechanisms as in \citeA{marshall2017} -- and a crucial aspect is the assumed linear scaling of eddy heat fluxes with eddy energy (in \citeA{ambaum2014}: heat fluxes scaling with the square of the eddy amplitude). The resulting system is a non-linear oscillator, whose equilibrium response exhibits the principles of eddy saturation and frictional control \cite{novak2018} (with the latter termed ``dissipative control'' in \citeA{novak2018}). A two-dimensional dynamical system for the Antarctic Slope Current has recently been discussed in \citeA{ong2024}, based upon the the action of eddies on a constant density slope in a two-layer model, with the GEOMETRIC energy scaling. These are the same key physical principles as in the \citeA{ambaum2014} model, and lead to the same two-dimensional dynamical system (up to the definitions of constants, and reached from equations (6)--(7) in \citeA{ong2024} by deriving an equation for $\sqrt{APE}$). A key result from the \citeA{ambaum2014} and \citeA{ong2024} models is the prediction of an intrinsic oscillatory time scale.

A two-dimensional dynamical system for Southern Ocean mean available potential energy and eddy kinetic energy is also derived in \citeA{sinha2016}, and the key physical principles are again the same: erosion of a measure of the mean flow by baroclinic instability, feeding eddy energy. Subject to a choice of parameters, in particular choosing an eddy energy generation term linear in eddy energy and a linear damping (respectively $\alpha = 1$ and $\beta = 1$ in the notation of \citeA{sinha2016}) the \citeA{ambaum2014} dynamical system is again obtained up to the definitions of constants. However the \citeA{sinha2016} system includes an equation for mean available potential energy, rather than baroclinicity, and so has different physical behaviour.

\citeA{novak2017} note the relationship of the \citeA{ambaum2014} model to the earlier low-dimension models for the atmosphere described in \citeA{lorenz1984} and \citeA{thompson1987}. The \citeA{thompson1987} model is formed as a reduced order model for a two-level quasigeostrophic model on a $\beta$-plane. A version of this model is analysed in \citeA{kobras2022}, yielding a six dimensional system where eddy saturation and frictional/dissipative control mechanisms are discussed. This is further extended in \citeA{kobras2024} in a more complicated eight dimensional model, incorporating eddy tilt effects, where partial eddy saturation behaviour can be exhibited depending on the regime.

The appearance of the same dynamical equations, emerging based on the same physical principles, motivates a more detailed study of the consequences of these principles for the dynamics of the Antarctic Circumpolar Current. In this article we return to a model of the Antarctic Circumpolar Current as in \citeA{marshall2017}, and consider a simple dynamical extension. The model is specifically derived using a reduced order density equation with a single dynamical degree of freedom setting the meridional mean density gradient, combined with a version of the GEOMETRIC eddy energy budget. The model as derived here is not dissimilar to the zonally averaged planetary geostrophic channel model in \citeA{mak2017}, but is constructed using the further simplifications of a fixed linear stratification, and a time varying linear lateral density gradient \cite<cf.>[]{thompson2016}. Specifically the mean thermal wind transport equation is arrived at using a Galerkin discretization with a single spatial degree of freedom which sets the lateral mean density gradient. Since the model derived here uses the physical principles of \citeA{ambaum2014}, the same dynamical system is once again reached (up to the definitions of constants). As a result we here obtain a simple two-dimensional model for the Antarctic Circumpolar Current which exhibits eddy saturation and frictional control in its equilibrium response, and which also predicts an intrinsic oscillatory time scale when perturbed from equilibrium.

Sensitivities to more rapid fluctuations on top of the mean surface wind stress may also be important. Eddy saturation and frictional control relate to the baroclinic processes, which operate on time scales longer than barotropic responses that are known to operate in the Southern Ocean \cite<e.g.,>[]{hughes1999, sura2003, olbers2007}. The faster time responses are usually attributed to the short term sub-annual fluctuations in the surface wind forcing over the Southern Ocean, and sometimes modelled as stochastic \cite<e.g.,>[]{sura2003}. It is of interest to investigate how the transport and eddy energy responds to more rapid fluctuations, and whether such fluctuations have consequences for the eddy saturation and frictional control phenomena. This is investigated in this article by considering a simple stochastic extension to the two-dimensional dynamical system, and it is found that while the mean thermal wind transport is relatively less sensitive to the stochastic forcing, the eddy energy exhibits more significant excursions away from equilibrium.

The article proceeds as follows. In section \ref{sect:scalings} the scalings of the \citeA{marshall2017} description are summarised. In section \ref{sect:formulation} this is extended via the formulation of a two component dynamical system, which has a Hamiltonian limit governed by the non-linear oscillator dynamics of \citeA{ambaum2014} and whose fixed-point exhibits the scalings of the \citeA{marshall2017} model in this limit. The response of this system to varying wind forcing is considered in section \ref{sect:forcing}, considering first the basic case of the linear response to oscillatory wind forcing, before considering the response to stochastic wind forcing. The paper concludes in section \ref{sect:conclusions}.


\section{Key scalings}\label{sect:scalings}

The principles described in \citeA{marshall2017} are briefly outlined. The discussion in \citeA{marshall2017} is phrased in terms of vertical momentum eddy stress. Here, and consistent with the formulation to follow, an equivalent viewpoint in terms of Eulerian mean and eddy-induced overturning is outlined. The outline here follows a basic scaling argument, and the scalings are justified more robustly, with dimensional factors restored, in section \ref{sect:formulation}.

A zonally periodic channel is considered, subject to a surface wind stress forcing with magnitude $\tau_0$. The mean zonal flow has magnitude $U$ near the surface and, as per the argument in \citeA{straub1993}, is assumed to be weak at depth.

Considering first the mean momentum balance the vertical shear leads, through thermal wind balance, to an Eulerian mean overturning, suggesting
\begin{linenomath*}\begin{equation}
  \textrm{Eulerian mean overturning} \sim \tau_0.
\end{equation}\end{linenomath*}
This Eulerian mean overturning is countered by an eddy-induced overturning associated with the baroclinically unstable vertical shear. If the (total) eddy energy is $E$ then, consistent with the GEOMETRIC scheme, this suggests the linear scaling
\begin{linenomath*}\begin{equation}
  \textrm{Eddy-induced overturning} \sim E.
\end{equation}\end{linenomath*}
Mean momentum balance therefore suggests
\begin{linenomath*}\begin{equation}
  E \sim \tau_0,
\end{equation}\end{linenomath*}
i.e. that the eddy energy increases with increasing surface wind stress magnitude.

Now, considering the eddy energy balance, baroclinic instability leads to generation of eddy energy
\begin{linenomath*}\begin{equation}
  \textrm{Eddy energy generation} \sim \textrm{Vertical shear} \times \textrm{Eddy energy} \sim U E.
\end{equation}\end{linenomath*}
which uses a GEOMETRIC form for the eddy form stress \cite{marshall2012,marshall2017}. Assuming that eddy energy is dissipated linearly at a rate $\lambda$, i.e.,
\begin{linenomath*}\begin{equation}
  \textrm{Eddy energy dissipation} \sim \lambda E,
\end{equation}\end{linenomath*}
eddy energy balance therefore suggests
\begin{linenomath*}\begin{equation}
  U \sim \lambda.
\end{equation}\end{linenomath*}
This is the ``frictional control'' mechanism described in \citeA{marshall2017}, with the mean flow magnitude increasing with increasing dissipation. Since $U$ is independent of $\tau_0$, these arguments imply eddy saturation.

These arguments, which follow \citeA{marshall2017}, imply that the momentum input $\tau_0$ sets the scale of the total eddy energy $E$, and the magnitude of eddy energy dissipation $\lambda$ sets the mean zonal momentum.
\section{Dynamical equations}\label{sect:formulation}


\subsection{Density profile}

A zonally periodic channel $\left( x, y, z \right) \in \left[ 0, L_x \right] \times \left[ -L / 2, L / 2 \right] \times \left[ -H, 0 \right]$ is considered with a linearly varying mean density profile defined (up to the addition of a constant) to be
\begin{linenomath*}\begin{equation}\label{eqn:density_1}
  \rho \left( y, z, t \right) = -\frac{\rho_0}{g} m \left( t \right) y - \frac{\rho_0}{g} N_0^2 z.
\end{equation}\end{linenomath*}
Here $\rho_0$ is a constant reference density, $g$ the magnitude of the gravitational acceleration, and $N_0$ the buoyancy frequency, all assumed to be constant. A time-dependent function $m \left( t \right)$ sets the horizontal density gradient. The domain integrated density is constant for all $m \left( t \right)$, which ensures conservation of the integrated density in the reduced order model to follow. The zonal, meridional, and vertical coordinates are denoted $x$, $y$, and $z$ respectively, and $t$ is time.

Thermal wind balance on an $f$-plane leads to a mean thermal wind transport
\begin{linenomath*}\begin{equation}\label{eqn:T_m}
  T \left( t \right) = -\frac{1}{2} L H^2 \frac{1}{f_0} m \left( t \right),
\end{equation}\end{linenomath*}
where $f_0$ is the Coriolis parameter. Hence the mean density may be written
\begin{linenomath*}\begin{equation}\label{eqn:density_2}
  \rho \left( y, z, t \right) = 2 \frac{1}{L H^2} \frac{f_0 \rho_0}{g} T \left( t \right) y - \frac{\rho_0}{g} N_0^2 z.
\end{equation}\end{linenomath*}


\subsection{Mean thermal wind transport}

Consider the mean density equation
\begin{linenomath*}\begin{equation}\label{eqn:mean_density}
  \frac{\partial \rho}{\partial t} + \frac{\partial}{\partial y} \left[ \left( v + v^* \right) \rho \right] + \frac{\partial}{\partial z} \left[ \left( w + w^* \right) \rho \right] = 0,
\end{equation}\end{linenomath*}
where $v$ and $w$ are the meridional and vertical components of the mean velocity, and $v^*$ and $w^*$ are corresponding components of an eddy-induced transport velocity. We now derive a reduced order model using a Galerkin spatial discretisation of the density equation \eqref{eqn:mean_density}, with the time-dependent part of the discrete density having a single spatial degree of freedom setting the lateral density gradient as per equation \eqref{eqn:density_2}.

Multiplying by $g y$ and integrating leads to
\begin{linenomath*}\begin{equation}\label{eqn:rho_y_integral}
  \frac{\mathrm{d}}{\mathrm{d} t} \int_{-L/2}^{L/2} \int_{-H}^0 \rho g y \, \mathrm{d}z \, \mathrm{d}y
    = \int_{-L/2}^{L/2} \int_{-H}^0 \left( v + v^* \right) \rho g \, \mathrm{d}z \, \mathrm{d}y,
\end{equation}\end{linenomath*}
where no-normal-flow boundary conditions are assumed. For the density profile \eqref{eqn:density_2},
\begin{linenomath*}\begin{equation}\label{eqn:rho_t}
  \frac{\mathrm{d}}{\mathrm{d} t} \int_{-L/2}^{L/2} \int_{-H}^0 \rho g y \, \mathrm{d}z \, \mathrm{d}y
    = \frac{1}{6} \frac{L^2}{H} f_0 \rho_0 \frac{\mathrm{d} T}{\mathrm{d} t}.
\end{equation}\end{linenomath*}

The planetary geostrophic mean zonal momentum equation gives
\begin{linenomath*}\begin{equation}
  -f_0 v = F - D,
\end{equation}\end{linenomath*}
where $F$ and $D$ represent mean zonal momentum forcing and dissipation respectively. Assuming that the system is forced by a wind stress concentrated to a near surface region, balanced by a bottom stress of equal and opposite magnitude concentrated to a near base region, leads to
\begin{linenomath*}\begin{equation}
  -\int_{-L/2}^{L/2} \int_{-H}^0 f_0 v \rho \, \mathrm{d}z \, \mathrm{d}y
    = \frac{\tau_0}{\rho_0} \int_{-L/2}^{L/2} \int_{-H}^0 \left[ \delta(z) - \delta \left( z + H \right) \right] \rho \, \mathrm{d}z \, \mathrm{d}y,
\end{equation}\end{linenomath*}
where $\tau_0$ defines the surface wind stress. For the density profile \eqref{eqn:density_2} this gives
\begin{linenomath*}\begin{equation}\label{eqn:rho_mean_adv}
  \int_{-L/2}^{L/2} \int_{-H}^0 v \rho g \, \mathrm{d}z \, \mathrm{d}y
    = L H \frac{N_0^2}{f_0} \tau_0.
\end{equation}\end{linenomath*}

Writing the eddy-induced transport velocity in terms of the eddy transport stream function
\begin{linenomath*}\begin{equation}
  v^* = \frac{\partial \psi}{\partial z}
\end{equation}\end{linenomath*}
leads to
\begin{linenomath*}\begin{align}
  \int_{-L/2}^{L/2} \int_{-H}^0 v^* \rho g \mathrm{d}z \, \mathrm{d}y
    & = \int_{-L/2}^{L/2} \int_{-H}^0 \frac{\partial \psi}{\partial z} \rho g\, \mathrm{d}z \, \mathrm{d}y \nonumber \\
    & = -g \int_{-L/2}^{L/2} \int_{-H}^0 \psi \frac{\partial \rho}{\partial z}\, \mathrm{d}z \, \mathrm{d}y,
\end{align}\end{linenomath*}
where $\psi = 0$ on boundaries has been used. Applying the Gent--McWilliams parameterisation \cite{gent1990,gent1995} we have
\begin{linenomath*}\begin{equation}
  \psi = \kappa_\text{GM} \frac{\partial \rho}{\partial y} \left( \frac{\partial \rho}{\partial z} \right)^{-1}.
\end{equation}\end{linenomath*}
Being precise this expression is applied only on the interior, and $\psi = 0$ is applied on the boundaries. Equivalently in the following $\kappa_\text{GM}$ may be defined so that it falls rapidly to zero near the boundaries. Assuming that $\kappa_\text{GM}$ is spatially constant, and for the mean density profile \eqref{eqn:density_2}, we obtain
\begin{linenomath*}\begin{equation}\label{eqn:rho_eddy_adv}
  \int_{-L/2}^{L/2} \int_{-H}^0 v^* \rho g\, \mathrm{d}z \, \mathrm{d}y
    = -2 \frac{1}{H} f_0 \rho_0 \kappa_\text{GM} T.
\end{equation}\end{linenomath*}

Combining equations \eqref{eqn:rho_y_integral}, \eqref{eqn:rho_t}, \eqref{eqn:rho_mean_adv}, and \eqref{eqn:rho_eddy_adv} leads to an equation for the mean thermal wind transport
\begin{linenomath*}\begin{equation}\label{eqn:T}
    \frac{\mathrm{d} T}{\mathrm{d} t} =
    6 \frac{H^2}{L} \frac{N_0^2}{f_0^2} \frac{\tau_0}{\rho_0}
  - 12 \frac{1}{L^2} \kappa_\text{GM} T.
\end{equation}\end{linenomath*}
The two right-hand-side terms correspond to the competition between generation of mean thermal wind transport due to the wind-driven Eulerian overturning (first term), and erosion by baroclinic instability (second term). Alternatively this may be interpreted in terms of a balance between near surface wind and eddy stress,
\begin{linenomath*}\begin{equation}
    \frac{\mathrm{d} T}{\mathrm{d} t} =
    6 \frac{H^2}{L} \frac{N_0^2}{f_0^2} \frac{1}{\rho_0} \left[ \tau_0
  - 2 \rho_0 \frac{1}{L H^2} \frac{f_0^2}{N_0^2} \kappa_\text{GM} T \right],
\end{equation}\end{linenomath*}
where the second bracketed term is the vertical eddy stress.

The mean thermal wind transport equation arrived at is similar to the Southern Ocean component of the \citeA{gnanadesikan1999}-type model described in \citeA{allison2011}. In \citeA{allison2011} the evolution of a single iscopycnal is modelled, and the channel is coupled to an inter-hemispheric basin. Here a continuous linear stratification in a bounded channel is considered. Moreover in \citeA{allison2011} a fixed and constant value for the Gent--McWilliams coefficient $\kappa_\text{GM}$ is considered.


\subsection{GEOMETRIC and eddy energtics}

Eddy saturation is obtained at the steady-state of equation \eqref{eqn:T} if the magnitude of the Gent--McWilliams coefficient $\kappa_\text{GM}$ scales with the magnitude of the surface wind stress $\tau_0$. In the following such a scaling is obtained as an emergent property, through the definition of an appropriate form of energetically constrained Gent--McWilliams coefficient.

For the mean density profile \eqref{eqn:density_1} GEOMETRIC defines a Gent--McWilliams coefficient \cite{marshall2012}
\begin{linenomath*}\begin{equation}\label{eqn:geometric_gm}
  \kappa_\text{GM} = \alpha E \frac{N_0}{\left| m \right|},
\end{equation}\end{linenomath*}
where $\alpha$ is a non-dimensional constant which in the quasi-geostrophic limit has a maximum magnitude of one \cite{marshall2012}, and is here assumed to be positive, consistent with baroclinic instability of the mean flow. $E$ is the specific eddy energy, and the domain integrated total eddy energy is
\begin{linenomath*}\begin{equation}
  \mathcal{E} = L_x \int_{-L/2}^{L/2} \int_{-H}^0 \rho_0 E\, \mathrm{d}z \, \mathrm{d}y.
\end{equation}\end{linenomath*}
Assuming that both $\kappa_\text{GM}$ and $\alpha$ are spatially constant and using \eqref{eqn:T_m} yields (for $f_0 T$ negative)
\begin{linenomath*}\begin{equation}\label{eqn:geometric_kappa_gm}
  \kappa_\text{GM} = -\frac{1}{2} \alpha \frac{H}{L_x} \frac{N_0}{f_0} \frac{\mathcal{E}}{\rho_0 T}.
\end{equation}\end{linenomath*}

GEOMETRIC includes an equation for a depth-integrated parameterised total eddy energy budget. Here this is further simplified by considering a fully domain integrated budget \cite<see>[]{mak2017},
\begin{linenomath*}\begin{equation}\label{eqn:E}
  \frac{\mathrm{d} \mathcal{E}}{\mathrm{d} t}
    = - L_x \int_{-L/2}^{L/2} \int_{-H}^0 g \kappa_\text{GM} \left( \frac{\partial \rho}{\partial y} \right)^2 \left( \frac{\partial \rho}{\partial z} \right)^{-1}\, \mathrm{d}z \, \mathrm{d}y
      - \lambda \left( \frac{\mathcal{E}}{\mathcal{E}_1} \right)^{p - 1} \mathcal{E} + \lambda \mathcal{E}_0,
\end{equation}\end{linenomath*}
where $p > 0$ is some dissipation exponent. The first right-hand-side term represents the mean-to-eddy energy conversion, the second a dissipation of eddy energy, and the third a background source of eddy energy. $\mathcal{E}_1$ has dimensions of energy, and for $p = 1$ the constant $\lambda$ defines a linear damping rate for the eddy energy towards $\mathcal{E}_0$.


\subsection{Two-dimensional dynamical system}

Using the GEOMETRIC form of the Gent--McWilliams coefficient \eqref{eqn:geometric_kappa_gm} with spatially constant $\kappa_\text{GM}$ and $\alpha$, for the case where the product $f_0 T$ is negative and for the density profile \eqref{eqn:density_2}, equations \eqref{eqn:T} and \eqref{eqn:E} become
\begin{linenomath*}\begin{subequations}\label{eqn:system}
  \begin{align}
    \frac{\mathrm{d} T}{\mathrm{d} t} & =
      6 \frac{H^2}{L} \frac{N_0^2}{f_0^2} \left[
      \frac{\tau_0}{\rho_0}
    + \alpha \frac{f_0}{N_0} \frac{\mathcal{E}}{\rho_0 L_x L H} \right], \label{eqn:system_T} \\
    \frac{\mathrm{d} \mathcal{E}}{\mathrm{d} t} & =
      \left[ -2 \alpha \frac{1}{L H^2} \frac{f_0}{N_0} T - \lambda \left( \frac{\mathcal{E}}{\mathcal{E}_1} \right)^{p - 1} \right] \mathcal{E}
    + \lambda \mathcal{E}_0, \label{eqn:system_E}
  \end{align}
\end{subequations}\end{linenomath*}
yielding a two-dimensional system of ordinary differential equations for the mean thermal wind transport and total eddy energy.

If these equations are generalised to include positive $f_0 T$ (noting that this requires $\left| m \right| \rightarrow -\left| m \right|$ in \eqref{eqn:geometric_gm}) then this permits mean thermal wind transport reversal with baroclinic stability. This unphysical case is included as it is technically needed later, for example when stochastic forcing is added.


\subsection{Steady state, eddy saturation, and frictional control}

At steady-state we obtain mean thermal wind transport and eddy energy
\begin{linenomath*}\begin{subequations}\label{eq:fixpt}
  \begin{align}
    T_* & = -\frac{1}{2} \frac{1}{\alpha} L H^2 \frac{N_0}{f_0} \lambda \left[ \left( \frac{\mathcal{E}_*}{\mathcal{E}_1} \right)^{p - 1} - \frac{\mathcal{E}_0}{\mathcal{E}}_* \right], \\
    \mathcal{E}_* & = -\frac{1}{\alpha} L_x L H \frac{N_0}{f_0} \tau_0. \label{eqn:E_fp}
  \end{align}
\end{subequations}\end{linenomath*}
In particular for the case of linear eddy energy damping, $p = 1$, and with small background energy generation $\mathcal{E}_0$, the mean thermal wind transport is independent of wind stress $\tau_0$, and scales with the eddy energy dissipation $\lambda$. Note also that the steady-state mean thermal wind transport is set by a balance of terms in the eddy energy budget. Hence this steady-state response is consistent with the mechanism for eddy saturation and frictional control described in \citeA{marshall2017}.

For $p = 1$ and $\mathcal{E}_0 = 0$ the Southern Ocean relevant parameters as in Table~\ref{tab:parameters} lead to a steady state thermal wind transport of $T_* = 270$~Sv and a steady state Gent-McWilliams coefficient of $3000$~m$^2$s$^{-1}$. While the precise value for the transport is somewhat too high these have a reasonable order of magnitude.


\subsection{Non-dimensionalisation}

It is natural to use the steady-state to define a non-dimensionalisation, with non-dimensional mean thermal wind transport $\tilde{T}$ and eddy energy $\tilde{\mathcal{E}}$ defined
\begin{linenomath*}\begin{subequations}
  \begin{align}
    T & = \hat{T} \tilde{T}, \\
    \mathcal{E} & = \hat{\mathcal{E}} \tilde{\mathcal{E}},
  \end{align}
\end{subequations}\end{linenomath*}
with
\begin{linenomath*}\begin{subequations}
  \begin{align}
    \hat{T} & = -\frac{1}{2} \frac{1}{\alpha} L H^2 \frac{N_0}{f_0} \lambda, \\
    \hat{\mathcal{E}} & = \mathcal{E}_* = -\frac{1}{\alpha} L_x L H \frac{N_0}{f_0} \tau_0.
  \end{align}
\end{subequations}\end{linenomath*}
Non-dimensional eddy energy parameters are similarly defined $\tilde{\mathcal{E}}_0 = \mathcal{E}_0 /  \hat{\mathcal{E}}$, $\tilde{\mathcal{E}}_1 = \mathcal{E}_1 /  \hat{\mathcal{E}}$. The eddy energy dissipation parameter $\lambda$ defines an inverse time scale, leading to a non-dimensionalised time $\tilde{t} = \lambda t$. Considering the logarithm of the non-dimensionalised eddy energy $\tilde{M} = \ln \tilde{\mathcal{E}}$, then leads to
\begin{linenomath*}\begin{subequations}\label{eqn:system_nd}
  \begin{align}
    \frac{\mathrm{d} \tilde{T}}{\mathrm{d} \tilde{t}} & = -\tilde{\omega}_0^2 \left( e^{\tilde{M}} - 1 \right), \\
    \frac{\mathrm{d} \tilde{M}}{\mathrm{d} \tilde{t}} & =
      \left( \tilde{T} - 1 \right)
    + \left[ 1 - \left( \frac{e^{\tilde{M}}}{\tilde{\mathcal{E}}_1} \right)^{p - 1} \right]
    + \tilde{\mathcal{E}}_0 e^{-\tilde{M}},
  \end{align}
\end{subequations}\end{linenomath*}
where a non-dimensional squared angular frequency is defined
\begin{linenomath*}\begin{equation}
  \tilde{\omega}_0^2 = - 12 \alpha \frac{1}{L^2} \frac{N_0}{f_0} \frac{\tau_0}{\rho_0} \frac{1}{\lambda^2}.
\end{equation}\end{linenomath*}
For $p = 1$ and  $\tilde{\mathcal{E}}_0 = 0$ the \citeA{ambaum2014} model is now obtained. The case $p = 2$ and $\tilde{\mathcal{E}}_0 = 0$ is considered in \citeA{federer2021}.


\subsection{Hamiltonian structure}

Considering the case $p = 1$ and $\tilde{\mathcal{E}}_0 = 0$ leads to
\begin{linenomath*}\begin{subequations}\label{eqn:system_ham}
  \begin{align}
    \frac{\mathrm{d} \tilde{T}}{\mathrm{d} \tilde{t}} & = -\tilde{\omega}_0^2 \left( e^{\tilde{M}} - 1 \right), \\
    \frac{\mathrm{d} \tilde{M}}{\mathrm{d} \tilde{t}} & = \left( \tilde{T} - 1 \right).
  \end{align}
\end{subequations}\end{linenomath*}
This corresponds directly to equations (6) in \citeA{ambaum2014}, and (6)--(7) in \citeA{ong2024} (in the latter after deriving an equation for $\sqrt{APE}$).
The corresponding non-linear and non-dimensional ``stiffness'' parameter is
\begin{linenomath*}\begin{equation}
  \tilde{k} = \tilde{\omega}_0 \sqrt{\frac{e^{\tilde{M}} - 1}{\tilde{M}}} = \tilde{\omega}_0 \sqrt{\frac{\tilde{\mathcal{E}} - 1}{\ln \tilde{\mathcal{E}}}},
\end{equation}\end{linenomath*}
which indicates that the eddy energy is more rapidly restored to equilibrium at large energy. Correspondingly, away from equilbrium the system spends longer periods at weak eddy energies with mean thermal wind transport growing due to Eulerian mean overturning, interspersed with relatively shorter bursts of higher eddy energy with mean thermal wind transport falling via baroclinic instability.

Defining a Hamiltonian \cite<equivalent up to scaling to (9) of>[]{ambaum2014}
\begin{linenomath*}\begin{equation}\label{eqn:ham}
  \mathcal{H} \left( \tilde{T}, \tilde{M} \right) = \frac{1}{2} \left( \tilde{T} - 1 \right)^2 + \tilde{\omega}_0^2 \left( e^{\tilde{M}} - 1 - \tilde{M} \right),
\end{equation}\end{linenomath*}
equations \eqref{eqn:system_ham} are equivalent to
\begin{linenomath*}\begin{subequations}
  \begin{align}
    \frac{\mathrm{d} \tilde{T}}{\mathrm{d} \tilde{t}} & = - \frac{\partial \mathcal{H}}{\partial \tilde{M}}, \\
    \frac{\mathrm{d} \tilde{M}}{\mathrm{d} \tilde{t}} & = + \frac{\partial \mathcal{H}}{\partial \tilde{T}}.
  \end{align}
\end{subequations}\end{linenomath*}
Here the Hamiltonian is defined such that it is non-negative and vanishes only at the fixed-point. With this convention the mean thermal wind transport is the conjugate momentum.

\begin{figure}
\begin{center}
\includegraphics[width=0.48\textwidth]{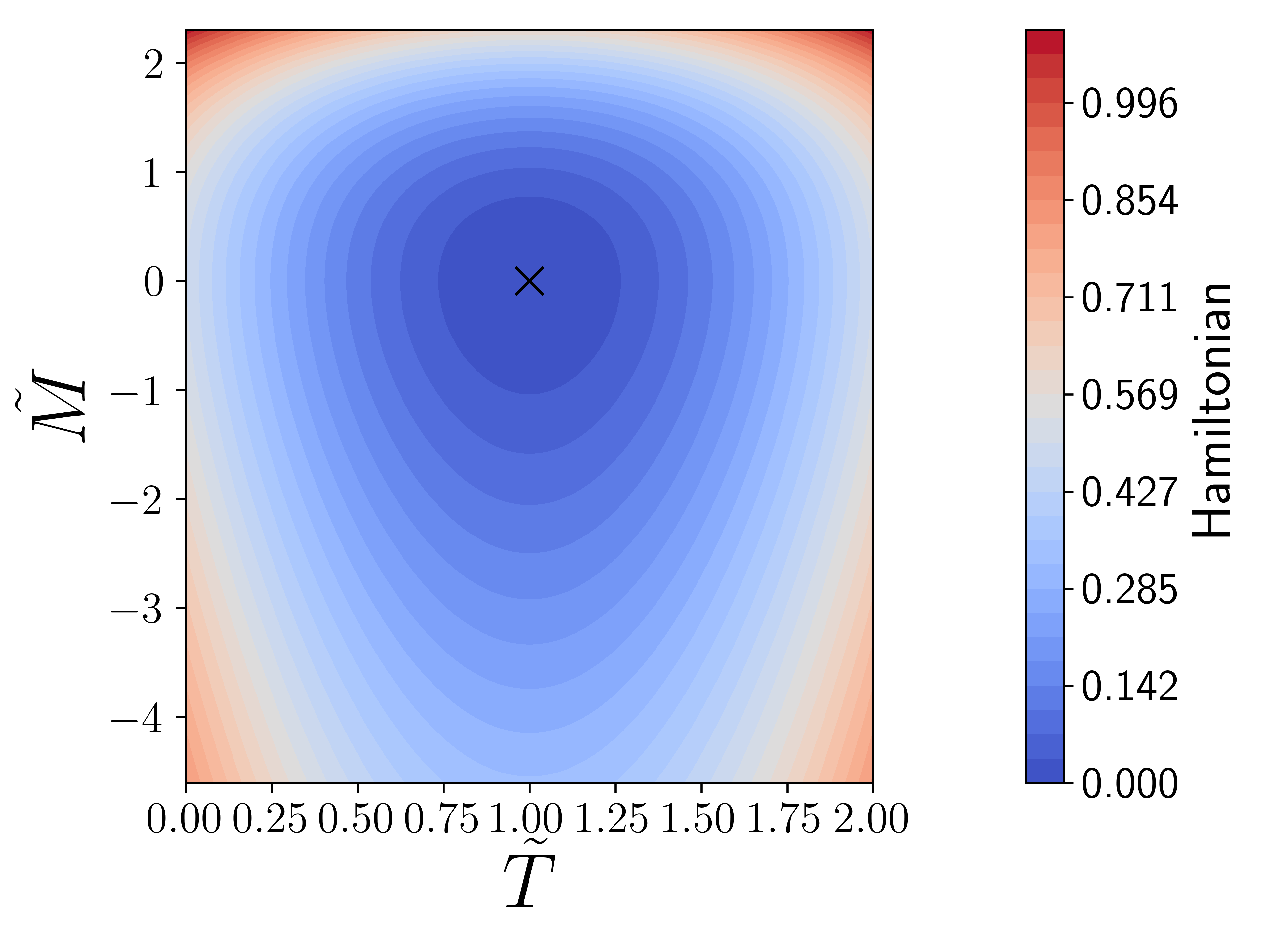}
~
\includegraphics[width=0.48\textwidth]{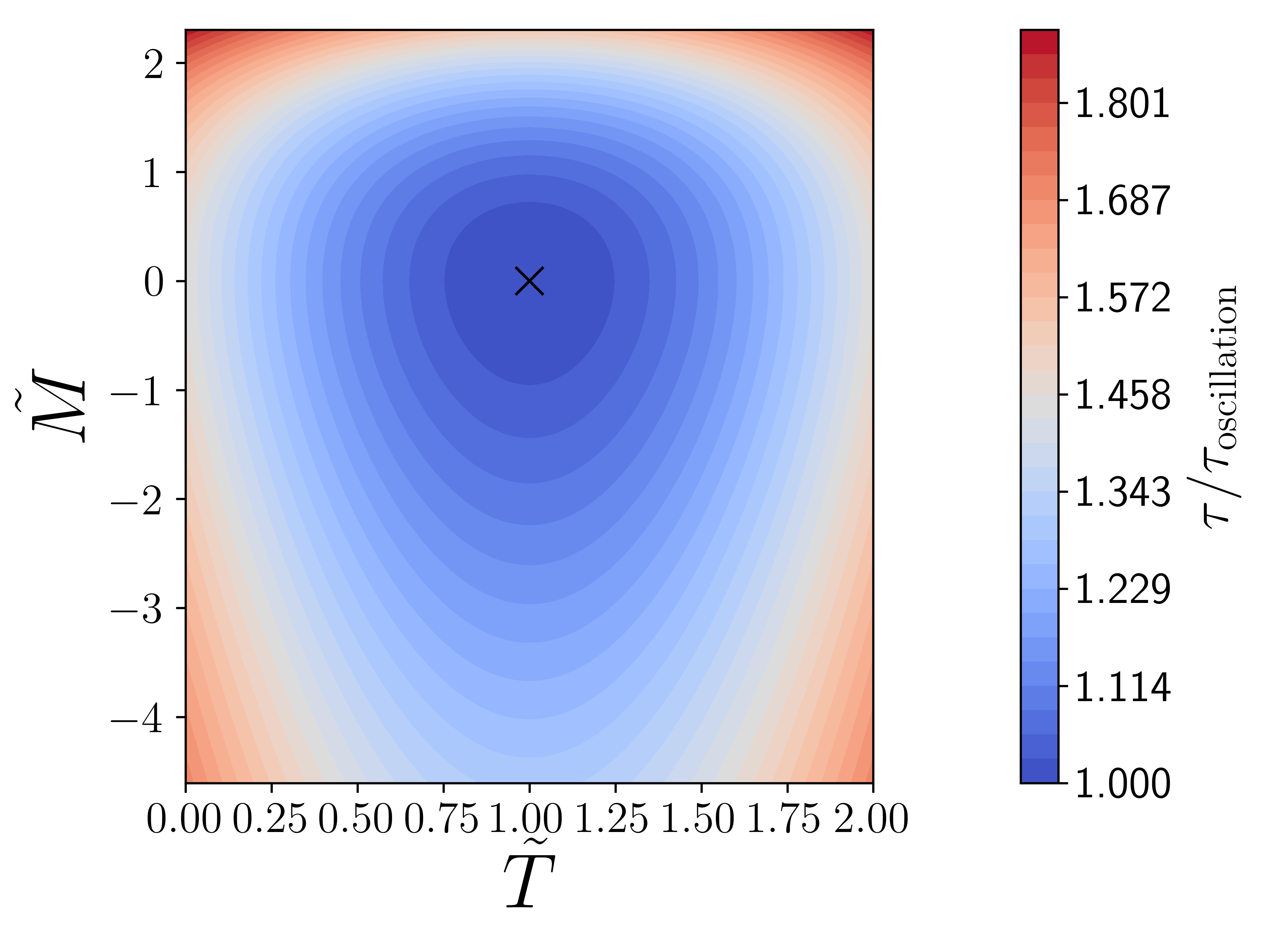}
\end{center}
\caption{Left: Hamiltonian \eqref{eqn:ham} for Southern Ocean relevant parameters as in Table~\ref{tab:parameters}, corresponding to $\tilde{\omega}_0^2 = 0.09$. The dynamics orbits on contours in an anti-clockwise sense. Right: Orbit time scales, computed using action-angle coordinators.}\label{fig:ham}
\end{figure}

\begin{figure}
\begin{center}
\begin{tabular}{ccccccc}
\includegraphics[width=0.19\textwidth]{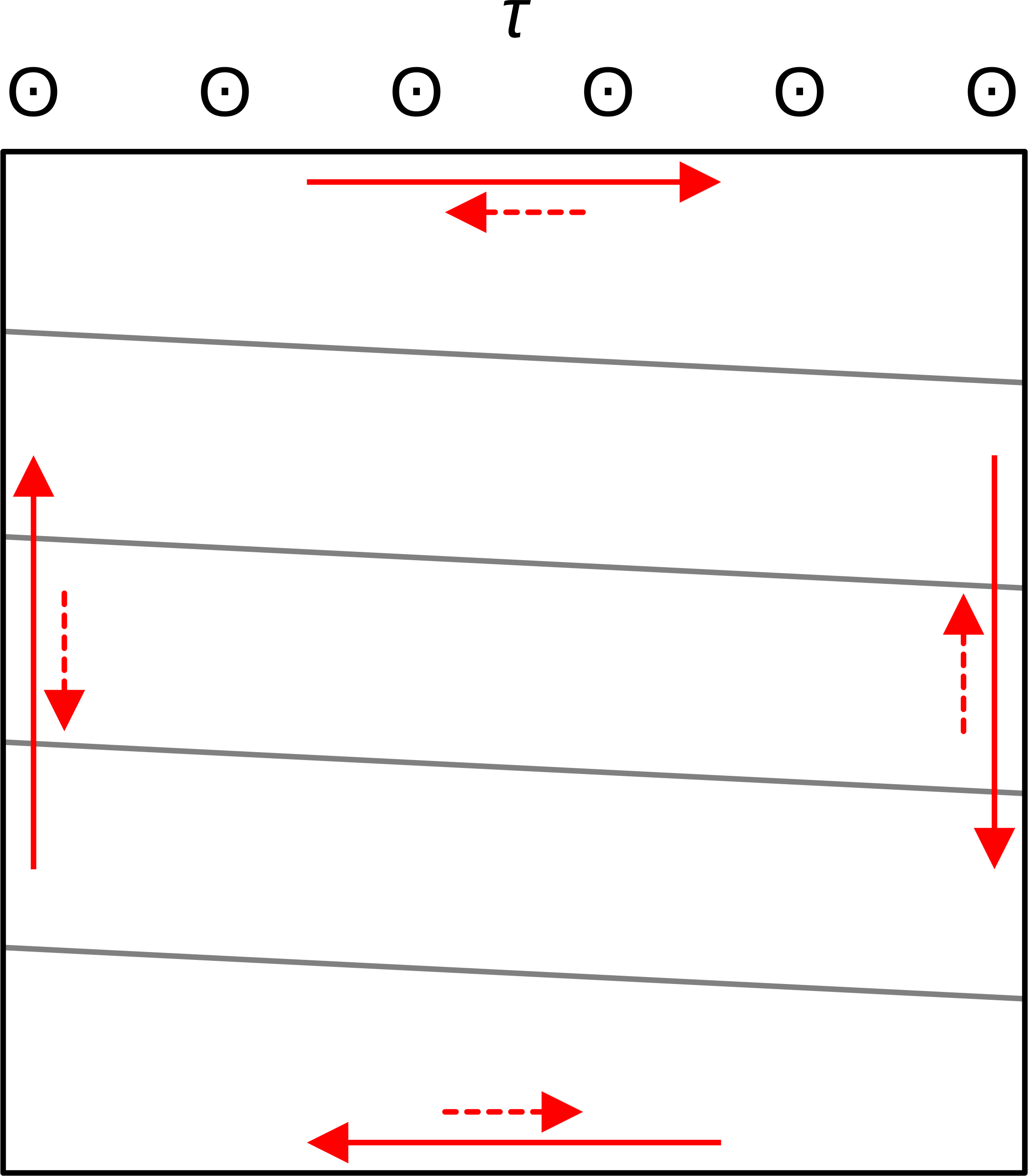}
& &
\includegraphics[width=0.19\textwidth]{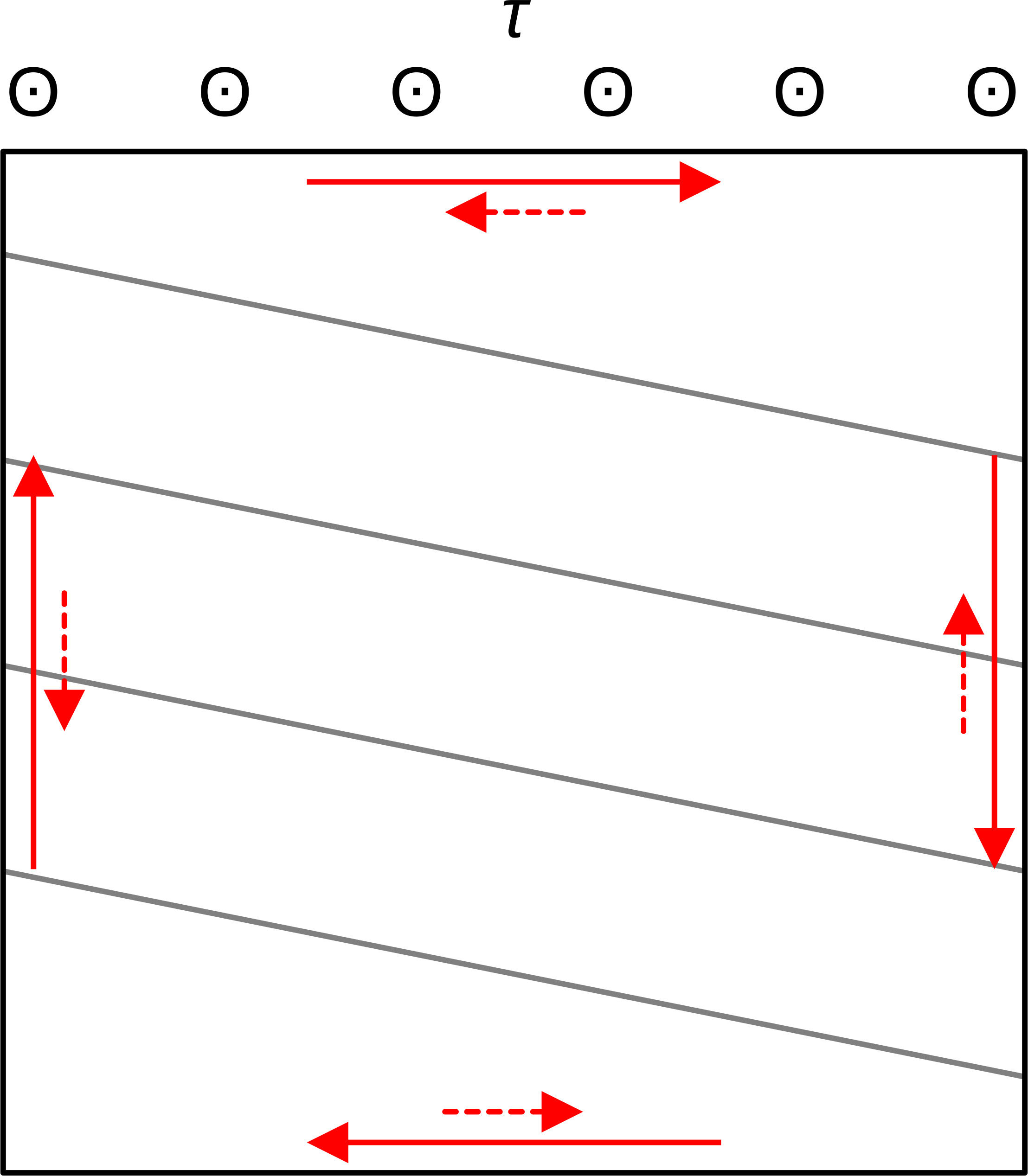}
& &
\includegraphics[width=0.19\textwidth]{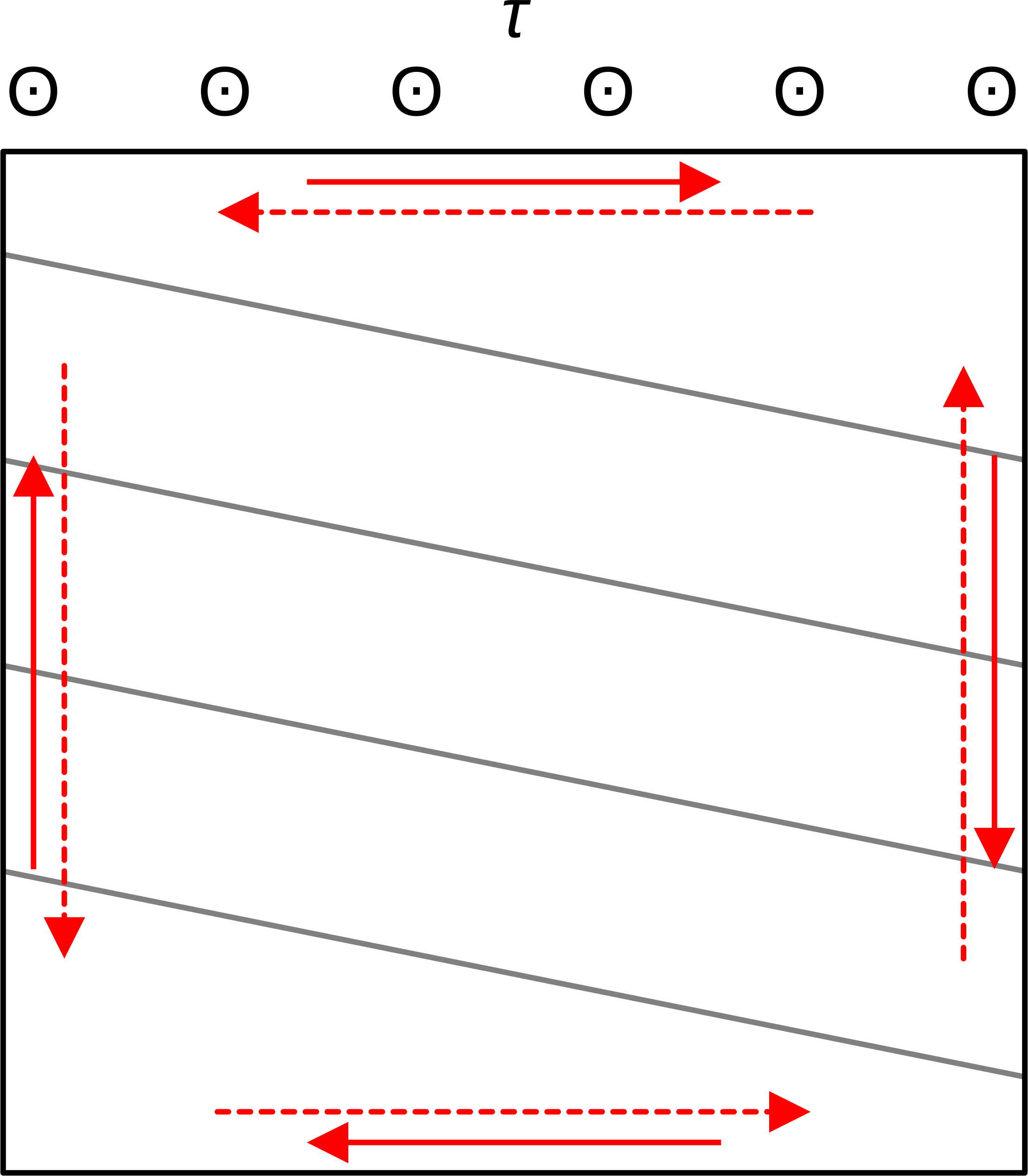}
& &
\includegraphics[width=0.19\textwidth]{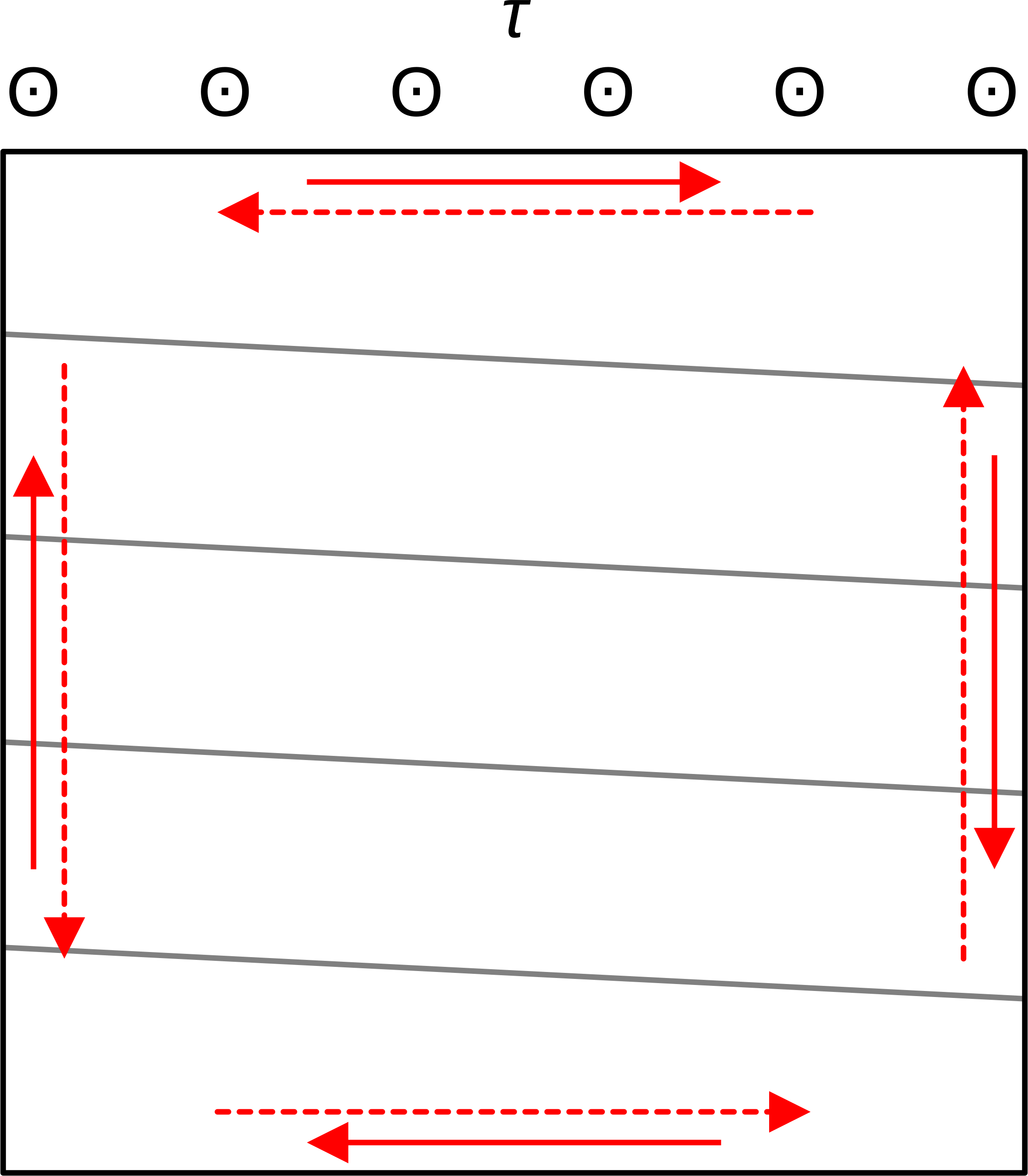} \\
A & & B & & C & & D
\end{tabular}
\end{center}
\caption{The mechanism of the non-linear oscillation. Density surfaces are indicated in grey, the Eulerian overturning circulation with solid arrows, and the eddy induced circulation with dashed arrows. A to B: The surface wind stress drives an Eulerian overturning, increasing lateral mean density gradients and hence the mean thermal wind transport. B to C: The increased mean thermal windshear leads to a growth in eddy energy via baroclinic instability. C to D: The eddy induced overturning overcomes the Eulerian overturning, eroding the lateral mean density gradients and hence the mean thermal wind transport. D to A: As the mean thermal windshear erodes, baroclinic instability weakens, and eddy energy decays. The same fundamental principles appear in Fig.~10 of \citeA{ong2024} in a model for the Antactic Slope Current.}\label{fig:schematic}
\end{figure}

Away from equilibrium the system orbits on contours of the Hamiltonian -- see Fig~\ref{fig:ham}. These orbits correspond to the mechanism shown schematically in Fig.~\ref{fig:schematic}, with the same fundamental principles appearing in Fig.~10 of \citeA{ong2024} in their model for the Antarctic Slope Current. Starting from low eddy energies and low mean thermal wind transport, the Eulerian mean overturning due to the wind stress is only weakly opposed, leading to a growth in the mean thermal windshear (lower part of the orbit in Fig.~\ref{fig:ham}). A sufficiently high shear drives an increase in the eddy energy via baroclinic instability (right part of the orbit) which, once the eddy energy is sufficiently high, counteracts the mean overturning and erodes the mean thermal windshear (upper part of the orbit). A sufficiently weakened mean thermal windshear, with correspondingly weakened eddy energy generation, is no longer able to support the eddy energy against dissipation, and so the eddy energy decays (left part of the orbit). The mechanism behind the oscillatory eddy energy response is reminiscent of the overshoot behaviour behind the oscillatory finite amplitude response in the inviscid cases of the two-layer model for baroclinic instability in \citeA{pedlosky1970}, and also of the ``charge and discharge'' regime described in \citeA{kobras2024}.

Differentiating the maximum non-dimensional mean thermal wind transport and eddy energy with respect to the Hamiltonian leads to
\begin{linenomath*}\begin{subequations}
  \begin{align}
    \frac{\mathrm{d} \tilde{T}_\text{max}}{\mathrm{d} \mathcal{H}} = \frac{1}{\tilde{T}_\text{max} - 1}, \\
    \frac{\mathrm{d} \tilde{\mathcal{E}}_\text{max}}{\mathrm{d} \mathcal{H}} = \frac{1}{\tilde{\omega}_0^2} \frac{\tilde{\mathcal{E}}_\text{max}}{\tilde{\mathcal{E}}_\text{max} - 1}.
  \end{align}
\end{subequations}\end{linenomath*}
Note the extra factor of $1 / \tilde{\omega}_0^2$ which appears in the latter expression, for the eddy energy, but not in the former, for the mean thermal wind transport. For Southern Ocean relevant parameters listed in Table~\ref{tab:parameters}, the parameter $\tilde{\omega}_0^2 = 0.09$ is relatively small, amplifying the effect on the eddy energetics when moving to more distant Hamiltonian contours. That is, perturbed dynamics can exhibit excursions to relatively large eddy energy.

\begin{table}
\begin{center}
\begin{tabular}{c|c|c}
Parameter & Symbol & Value \\
\hline
Wind stress & $\tau_0$ & $0.1$~N~m${}^{-2}$ \\
Buoyancy frequency & $N_0$ & $-30 f_0$ \\
Density & $\rho_0$ & $10^3$~kg~m${}^{-3}$ \\
Meridional domain size & $L$ & $2000$~km \\
Depth & H & 3~km \\
\hline
GEOMETRIC eddy efficiency parameter & $\alpha$ & $0.1$ \\
Eddy energy dissipation rate & $\lambda$ & $10^{-7}$~s${}^{-1}$
\end{tabular}
\caption{Southern Ocean relevant parameters. See e.g. \citeA{mak2022b} for eddy energy dissipation rate estimates. These define a non-dimensional squared angular frequency $\tilde{\omega}_0^2 = 0.09$, and correspond to an oscillatory time scale of $\sim 2 \pi / \left( \tilde{\omega}_0 \lambda \right) = 6.6$~years.}\label{tab:parameters}
\end{center}
\end{table}


\section{Response to forcing}\label{sect:forcing}


\subsection{Time scales}

At the fixed point \eqref{eq:fixpt} the Jacobian of the system \eqref{eqn:system} has eigenvalues
\begin{linenomath*}\begin{equation}
  \Lambda_\pm = -\frac{1}{\tau_\text{decay}} \pm i \frac{2 \pi}{\tau_\text{oscillation}},
\end{equation}\end{linenomath*}
with
\begin{linenomath*}\begin{subequations}
  \begin{align}
    \tau_\text{decay} & = \frac{2}{\frac{\mathcal{E}_0}{\mathcal{E}_*}
                              + \left( p - 1 \right) \left( \frac{\mathcal{E}_*}{\mathcal{E}_1} \right)^{p - 1}} \frac{1}{\lambda}, \\
    \tau_\text{oscillation} & = \frac{2 \pi}{\sqrt{\omega_0^2 - \frac{1}{\tau_\text{decay}^2}}},
  \end{align}
\end{subequations}\end{linenomath*}
where $\omega_0 = \tilde{\omega}_0 \lambda$.

Assuming $\tau_\text{decay}$ is sufficiently long so that $\tau_\text{oscillation}$ is real, if either $\mathcal{E}_0 > 0$ and $p \ge 1$, or $\mathcal{E}_0 \ge 0$ and $p > 1$, then the fixed-point is a stable focus with trajectories spiraling anti-clockwise in ($\tilde{T}, \tilde{M}$) phase space towards the fixed point \eqref{eq:fixpt} with linearised decay time scale $\tau_\text{decay}$.

Since the equilibrium eddy energy $\mathcal{E}_*$ scales linearly with the wind stress magnitude, for $\mathcal{E}_0 > 0$ and $p = 1$ the decay time scale scales linearly with the wind stress magnitude, while for $\mathcal{E}_0 = 0$ and $p > 1$ the decay time scale decreases with increasing wind stress magnitude.

\begin{figure}
\begin{center}
\includegraphics[width=0.8\textwidth]{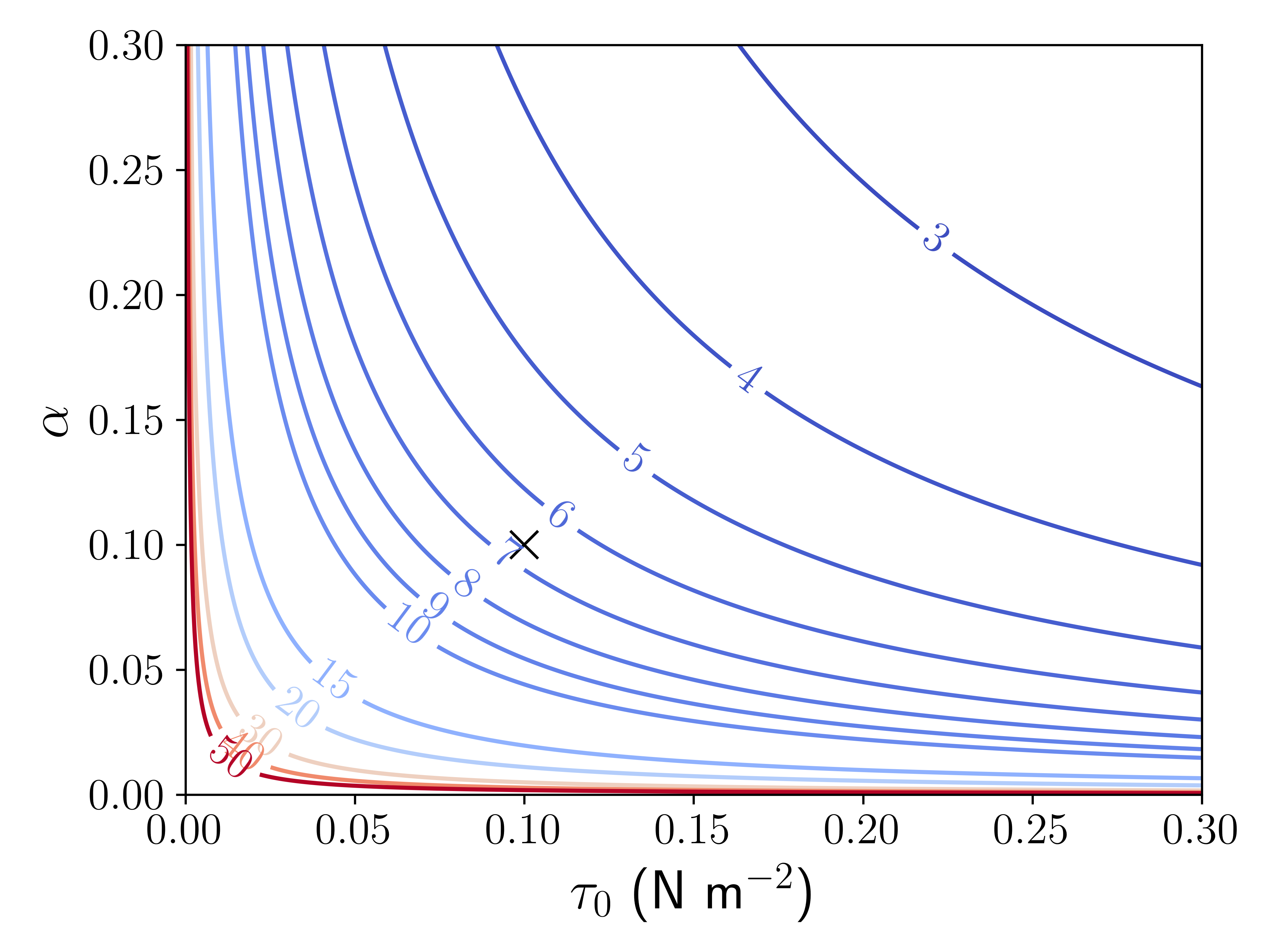}
\end{center}
\caption{The oscillatory time scale, in years, given by \eqref{eqn:t_oscillator} for different values of $\tau_0$ and $\alpha$. The black cross corresponds to $\tau_0 = 0.1$~N~m${}^{-2}$ and  $\alpha = 0.1$. Other parameters are as in Table~\ref{tab:parameters}.}\label{fig:oscillatory}
\end{figure}

For a long decay time scale there is a linearised oscillatory time scale
\begin{linenomath*}\begin{equation}\label{eqn:t_oscillator}
  \tau_\text{oscillation}
    = \frac{2 \pi}{\omega_0}
    = \frac{\pi}{\sqrt{3}} \sqrt{-\frac{1}{\alpha} L^2 \frac{f_0}{N_0} \frac{\rho_0}{\tau_0}},
\end{equation}\end{linenomath*}
which is $1 / \sqrt{2 \alpha}$ times the oscillatory time scale in \citeA{ong2024}. To see the mechanisms setting the angular frequency $\omega_0$, and hence the time scale $\tau_\text{oscillation}$, note that the dynamical system \eqref{eqn:system} can be re-written, for the case $p = 1$ and $\mathcal{E}_0 = 0$,
\begin{linenomath*}\begin{subequations}
  \begin{align}
    \frac{\mathrm{d} S}{\mathrm{d} t} & =
      12 \frac{1}{L^2} \frac{N_0^2}{f_0^2} \left[
      \frac{\tau_0}{\rho_0}
    + \alpha \frac{f_0}{N_0} \frac{\mathcal{E}}{\rho_0 L_x L H} \right], \label{eqn:system_S} \\
    \frac{\mathrm{d} \mathcal{E}}{\mathrm{d} t} & =
      -\alpha \frac{f_0}{N_0} \mathcal{E} S - \lambda \mathcal{E}, \label{eqn:system_S_E}
  \end{align}
\end{subequations}\end{linenomath*}
where $S = 2 T / \left( L H^2 \right)$ is the mean thermal windshear. Hence the time scale of the oscillatory response is set by two effects: the production of mean vertical shear due to the wind forced Eulerian overturning, and the efficiency of eddy energy generation for a given eddy energy and mean vertical shear,
\begin{linenomath*}\begin{equation}
  \omega_0^2 = \underbrace{12 \frac{1}{L^2} \frac{N_0^2}{f_0^2} \frac{\tau_0}{\rho_0}}_\text{Mean shear production due to wind} \times \underbrace{-\alpha \frac{f_0}{N_0}}_\text{Energy conversion efficiency}.
\end{equation}\end{linenomath*}
For the Southern Ocean relevant parameters listed in Table~\ref{tab:parameters} the oscillatory time scale is $\tau_\text{oscillation} = 6.6$~years. Note that there is only a weak square root dependence on the key parameters, and hence a decadal oscillatory time scale is obtained for a broad range of plausible parameters; see Fig.~\ref{fig:oscillatory}.

The oscillatory time scale \eqref{eqn:t_oscillator} scales inversely with the square root of the wind stress forcing $\tau_0$ and also inversely with the square root of the efficiency parameter $\alpha$, but is independent of the eddy energy dissipation rate $\lambda$. Changing the eddy energy dissipation rate changes the critical mean thermal windshear at which baroclinic instability is sufficiently strong to overcome the linear energy dissipation -- and hence sets the mean thermal wind transport at steady state -- but the oscillatory time scale is set by the wind driven Eulerian mean overturning and the generation of eddy energy by baroclinic instability.

For the case $p = 1$ and $\mathcal{E}_0 = 0$ the Hamiltonian structure allows the non-linear orbit time scale, away from equilibrium, to be computed using action-angle coordinates. Time scales, for the Southern Ocean relevant parameters as in Table~\ref{tab:parameters}, are shown in Fig.~\ref{fig:ham}. These time scales are computed using fourth order finite differencing of the Hamiltonian contour areas, with the Hamiltonian contour areas computed using a method based on Simpson's rule. Even with very large perturbations away from the fixed-point, associated with variations in mean thermal wind transport of order 100\%, the time scale retains the same order of magnitude.


\subsection{Oscillatory forcing}

Adding an oscillatory component to the wind stress
\begin{linenomath*}\begin{equation}
  \tau_0 \rightarrow \tau_0 \left[ 1 + \epsilon \sin \left( \omega t \right) \right],
\end{equation}\end{linenomath*}
for some $\omega > 0$, if we consider the linearised system for the Hamiltonian case $p = 1$ and $\mathcal{E}_0 = 0$, then the system reduces to a standard forced linear oscillator, with a standard resonant response near the resonant angular frequency $\omega_0$, noting that here the logarithm of the non-dimensionalised eddy energy plays the role of the ``displacement'', and the non-dimensionalised transport perturbation the role of the ``momentum''.

Of interest is to note that, for high frequencies, $\omega > \omega_0$, in the linearised case with $p = 1$ and $\mathcal{E}_0 = 0$ the mean thermal wind transport response lags the wind forcing by $\pi / 2$, and the eddy energy response is out of phase with the wind forcing. For low frequencies, $\omega < \omega_0$, the mean thermal wind transport response leads the wind forcing by $\pi / 2$, and the eddy energy response is in phase with the wind forcing.

Numerical simulations of the non-linear \citeA{ambaum2014} system with oscillatory forcing are described in \citeA{federer2021}.


\subsection{Stochastic forcing}

The problem can be generalised to the case of stochastic forcing by instead considering
\begin{linenomath*}\begin{equation}
  \tau_0 \rightarrow \tau_0 \left[ 1 + X_t \right],
\end{equation}\end{linenomath*}
where $X_t$ is a stochastic process. Here stochastic processes depending on $t$ or $\tilde{t}$ are denoted using subscripts.

If $X_t$ is an Ornstein-Uhlenbeck process then in an appropriate short decorrelation time scale limit \cite<see>[section 5.1]{pavliotis2014} we now obtain the stochastic differential equations
\begin{linenomath*}\begin{subequations}\label{eqn:system_sde}
  \begin{align}
    \mathrm{d}T_t & =
      6 \frac{H^2}{L} \frac{N_0^2}{f_0^2} \left[
      \frac{\tau_0}{\rho_0}
    + \alpha \frac{f_0}{N_0} \frac{\mathcal{E}_t}{\rho_0 L_x L H} \right] \mathrm{d}t
    + 6 \frac{H^2}{L} \frac{N_0^2}{f_0^2} \sqrt{\frac{2 \mathbb{V} \left( \frac{\tau}{\rho_0} \right)}{\gamma}} \mathrm{d}W_t, \\
    \mathrm{d}\mathcal{E}_t & =
      \left[ -2 \alpha \frac{1}{L H^2} \frac{f_0}{N_0} T_t - \lambda \left( \frac{\mathcal{E}_t}{\mathcal{E}_1} \right)^{p - 1} \right] \mathcal{E}_t \mathrm{d}t
    + \lambda \mathcal{E}_0 \mathrm{d}t,
  \end{align}
\end{subequations}\end{linenomath*}
where $\mathbb{V} \left( \tau \right) = \rho_0^2 \mathbb{V} \left( \tau / \rho_0 \right)$ is the wind stress variance and $1 / \gamma$ the wind stress decorrelation time scale.

In the case $p = 1$ and $\mathcal{E}_0 = 0$ it follows, using It\^{o}'s formula, that
\begin{linenomath*}\begin{equation}
  \frac{\mathrm{d} \mathbb{E} \left( \mathcal{H}_{\tilde{t}} \right)}{\mathrm{d} \tilde{t}} = \frac{1}{2} \tilde{\sigma}^2.
\end{equation}\end{linenomath*}
where $\mathcal{H}$ is the Hamiltonian defined in \eqref{eqn:ham}, and where
\begin{linenomath*}\begin{equation}
  \tilde{\sigma} = \frac{\rho_0}{\tau_0} \tilde{\omega}_0^2 \sqrt{\lambda} \sqrt{\frac{2 \mathbb{V} \left( \frac{\tau}{\rho_0} \right)}{\gamma}}.
\end{equation}\end{linenomath*}
Hence by perturbing the Hamiltonian system by adding stochastic wind forcing the system drifts in expectation towards higher Hamiltonian values, associated with longer period orbits, further from the fixed point.

Considering $p \ne 1$ or $\mathcal{E}_0 \ne 0$ we instead obtain
\begin{linenomath*}\begin{equation}
  \frac{\mathrm{d} \mathbb{E} \left( \mathcal{H}_{\tilde{t}} \right)}{\mathrm{d} \tilde{t}} = \mathbb{E} \left[
      \tilde{\omega_0}^2
      \left( \tilde{\mathcal{E}}_{\tilde{t}} - 1 \right)
      \left( 1 - \left( \frac{\tilde{\mathcal{E}}_{\tilde{t}}}{\tilde{\mathcal{E}}_1} \right)^{p - 1}
        + \frac{\tilde{\mathcal{E}}_0}{\tilde{\mathcal{E}}_{\tilde{t}}} \right) \right]
      + \frac{1}{2} \tilde{\sigma}^2.
\end{equation}\end{linenomath*}
In particular for the case of linear damping, $p = 1$, and at steady-state, a simple expression for the expectation of one over the eddy energy is obtained,
\begin{linenomath*}\begin{equation}\label{eqn:one_over_E_nondim}
  \mathbb{E} \left( \frac{1}{\tilde{\mathcal{E}}_{\tilde{t}}} \right) = 1 +\frac{1}{2} \frac{1}{\tilde{\omega}_0^2} \frac{\tilde{\sigma}^2}{\tilde{\mathcal{E}}_0},
\end{equation}\end{linenomath*}
or restoring dimensions and expressing in terms of fundamental parameters
\begin{linenomath*}\begin{equation}\label{eqn:one_over_E_dim}
  \mathbb{E} \left( \frac{1}{\mathcal{E}_t} \right) = -\alpha \frac{1}{L_x L H} \frac{f_0}{N_0} \frac{1}{\tau_0} \left[
      1 + 12 \rho_0 \frac{L_x H}{L} \frac{N_0^2}{f_0^2} \frac{1}{\lambda} \frac{1}{\mathcal{E}_0} \frac{\mathbb{V} \left( \frac{\tau}{\rho_0} \right)}{\gamma} \right].
\end{equation}\end{linenomath*}
As the wind stress variance $\mathbb{V} \left( \tau \right)$ increases, the expectation of one over the eddy energy also increases. This counter-intuitive behaviour is consistent with the dynamics being forced away from the fixed point, since orbits further from the fixed point spend relatively longer at lower eddy energy. The magnitude of the departure of the expectation from the steady-state value is controlled by the competition between the magnitude of the stochastic forcing due to the wind stress, and the decay towards the steady-state due to the background eddy energy term. The second term in \eqref{eqn:one_over_E_nondim} expresses the non-dimensional relative strength of these two effects on the eddy energetics.

\begin{figure}
\begin{center}
\includegraphics[width=0.8\textwidth]{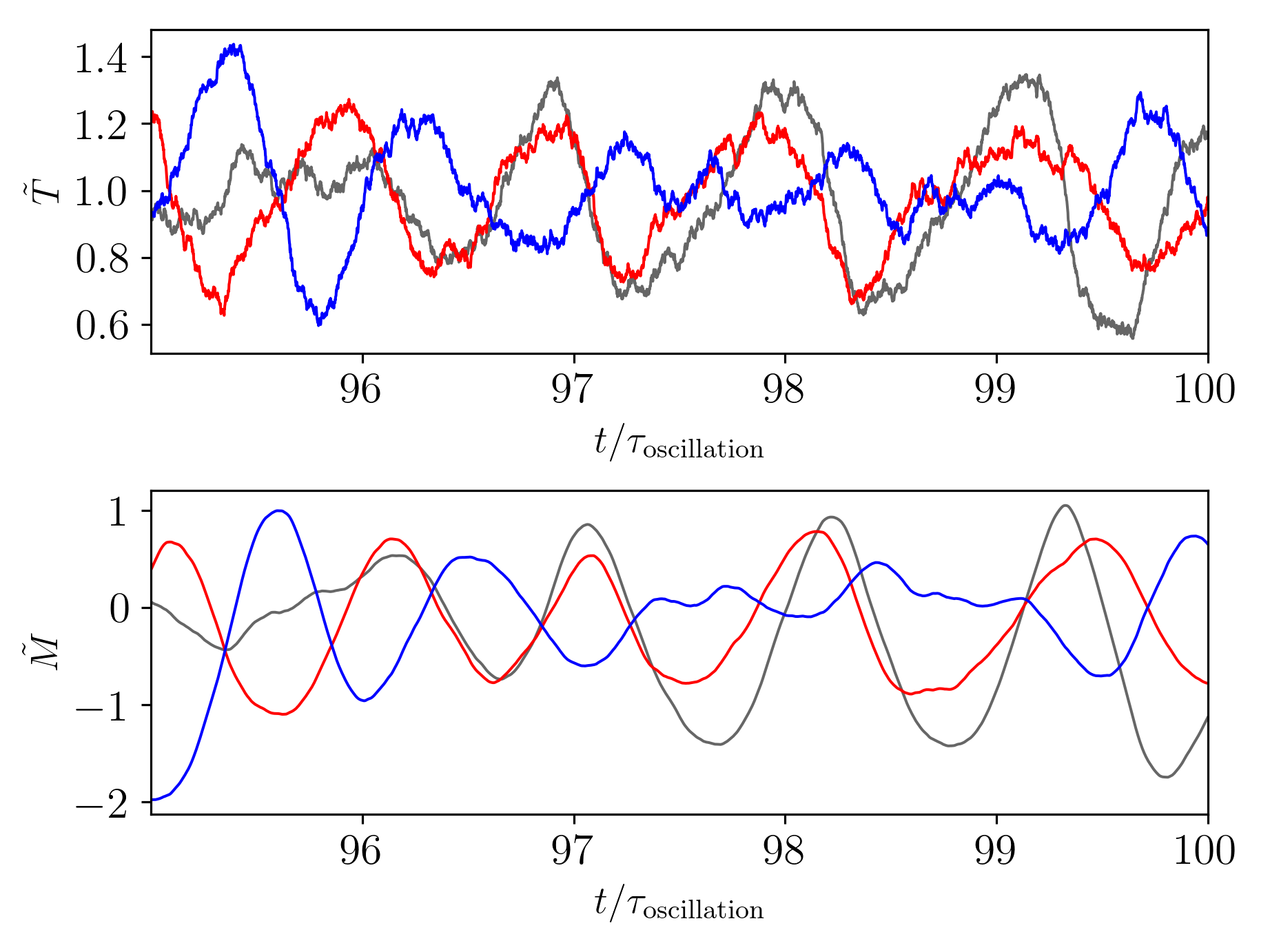}
\end{center}
\caption{Time series for three independent realisations of the stochastically forced system \eqref{eqn:system_sde}, with $\tilde{\omega}_0^2 = 0.09$, $\tilde{\mathcal{E}}_0 = 0.01$, and $\tilde{\sigma} = 0.04$.}\label{fig:T_M}
\end{figure}

\begin{figure}
\begin{center}
\includegraphics[width=0.8\textwidth]{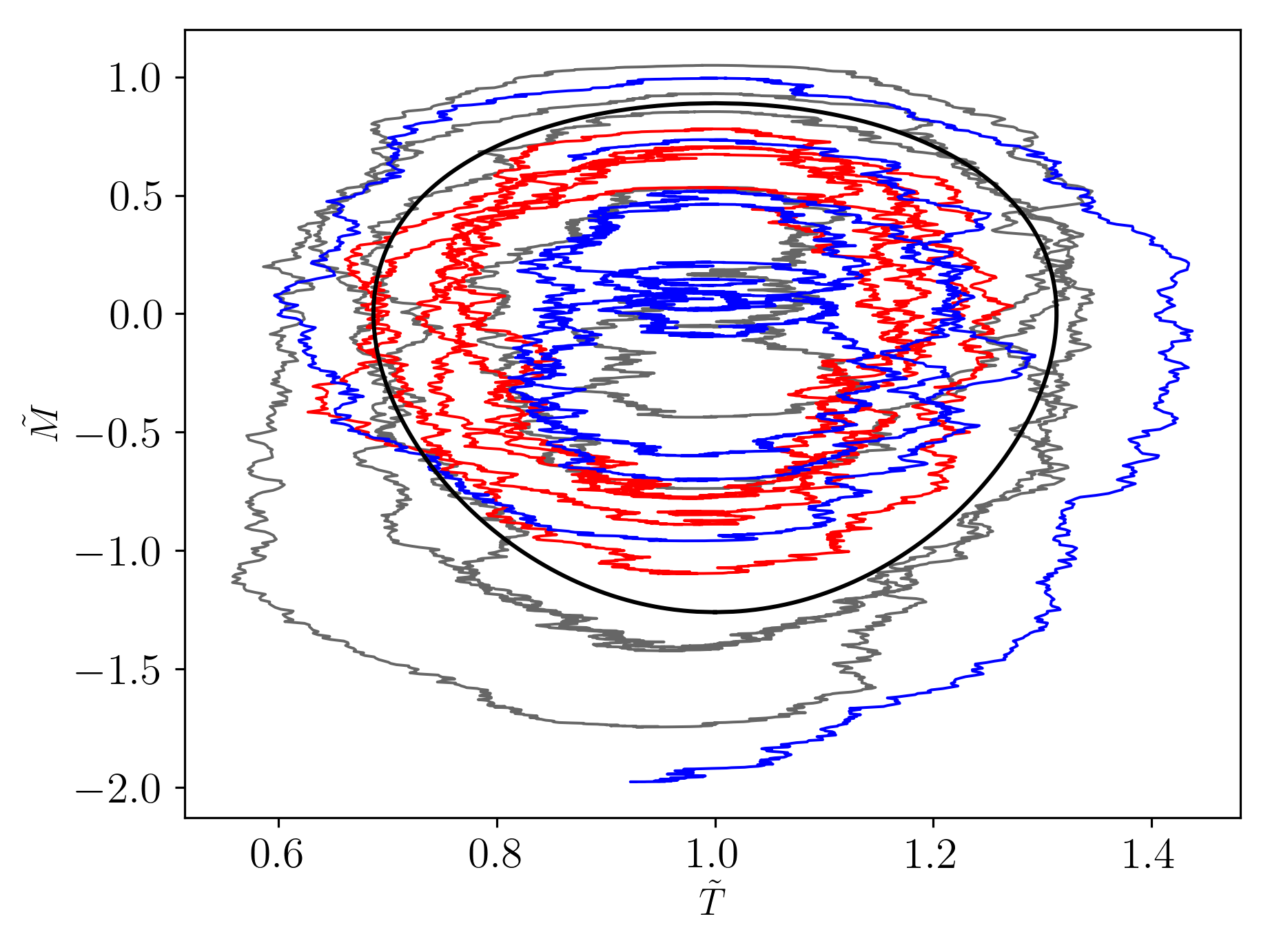}
\end{center}
\caption{Phase portrait for the realisations shown in Fig.~\ref{fig:T_M}. The black contour shows the Hamiltonian contour associated with the time average of the Hamiltonian from $t > 95 \tau_\text{oscillation}$ to $t \le 100 \tau_\text{oscillation}$, averaged over $10,000$ independent realisations (three of which are the realisations shown here).}\label{fig:T_M_phase}
\end{figure}

\begin{figure}
\begin{center}
\includegraphics[width=0.8\textwidth]{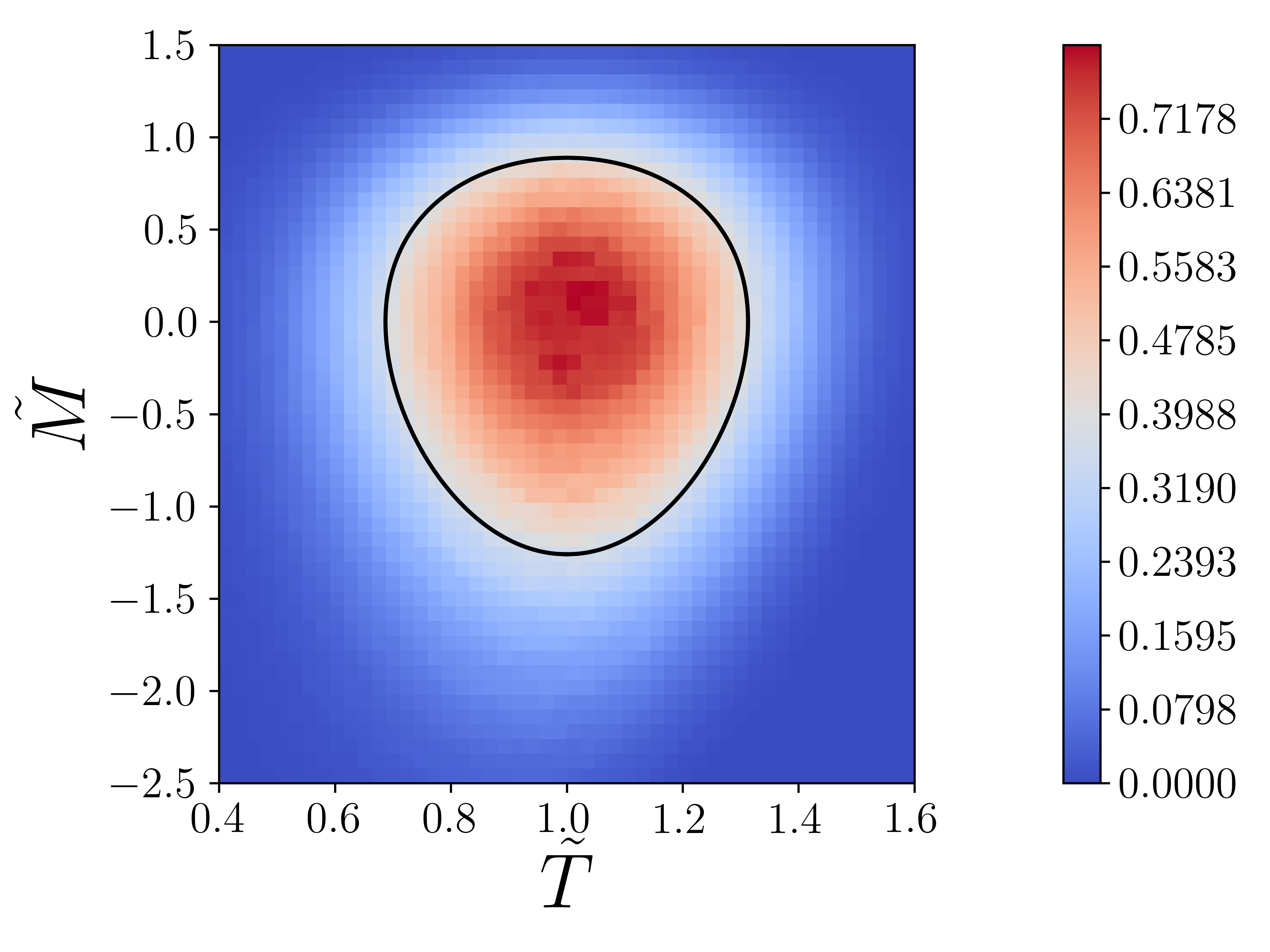}
\end{center}
\caption{Two-dimensional normalised histogram for $10,000$ independent realisations of the stochastically forced system \eqref{eqn:system_sde}, with $\tilde{\omega}_0^2 = 0.09$, $\tilde{\mathcal{E}}_0 = 0.01$, and $\tilde{\sigma} = 0.04$. The values obtained after each time step of size $10^{-3} \tau_\text{oscillation}$, for $t > 95 \tau_\text{oscillation}$ to $t \le 100 \tau_\text{oscillation}$ and for each realisation, are binned using a $50 \times 50$ array. $0.82$\% of values are outside the considered range of $\tilde{T}$ and $\tilde{M}$. The black contour is associated with the average of the Hamiltonian, and is as in Fig.~\ref{fig:T_M_phase}.}\label{fig:T_M_hist}
\end{figure}

Fig.~\ref{fig:T_M}, Fig.~\ref{fig:T_M_phase}, and Fig.~\ref{fig:T_M_hist} show the results from numerical solutions for the case $p = 1$. The background eddy energy is set to a weak value of $\mathcal{E}_0 / \mathcal{E}_* = 0.01$. The stochastic forcing is set via $\tilde{\sigma} = 0.04$, corresponding to a wind stress standard deviation of $128$\% of the mean wind stress for a wind stress decorrelation time scale of one week. Other parameters as in Table~\ref{tab:parameters}. The numerical calculations apply Strang splitting, with half Euler-Maruyama steps for the stochastic term before and after one full implicit mid-point rule step for the deterministic terms, and with a time step size of $10^{-3} \tau_\text{oscillation}$. The dynamics is perturbed away from the fixed point, although the effect on the characteristic oscillatory time scale is small -- the mean Hamiltonian contour, averaged from $t > 95 \tau_\text{oscillation}$ to $t \le 100 \tau_\text{oscillation}$ and over $10,000$ independent realisations, is associated with an orbit time scale 4.6\% larger than that at the fixed point. At these Southern Ocean relevant parameters, while the influence of the stochastic forcing on the mean thermal wind transport appears relatively more modest, the response of the eddy energy is substantial (noting that the \emph{logarithm} of the non-dimensional eddy energy is shown). Hence this simple model suggests that a more substantial response to stochastic wind forcing may be seen through its effect on the ocean turbulence, rather than through the behaviour of the mean. This effect arises not through direct driving of the turbulence, but instead through a perturbation of the system towards dynamics associated with larger excursions in eddy energy.


\section{Conclusions}\label{sect:conclusions}

This article describes a simple two component dynamical system for Southern Ocean mean zonal thermal wind transport, coupled to a parameterised mesoscale eddy field. The model captures the evolution of a planetary geostrophic channel with a linear mean density profile and constant stratification. The GEOMETRIC form of the Gent--McWilliams parameterision is used, and combined with a simple model for a domain integrated eddy energy budget.

In the case of linear damping and zero eddy energy background term the system is Hamiltonian, and is (up to the definition of constants) equivalent to the non-linear oscillator of \citeA{ambaum2014} and the system of \citeA{ong2024}. The steady-state mean thermal wind transport, set via the eddy energy balance, is insensitive to wind stress, capturing eddy saturation. The steady-state eddy energy, set via the balance between Eulerian and eddy-induced overturning, or equivalently via a balance of vertical stress, scales with the eddy energy damping rate, capturing the principle of frictional control. The system oscillates at a characteristic oscillatory time scale, set by the combined effect of mean shear generation by the Eulerian overturning and the efficiency of eddy energy generation for a given mean vertical shear and eddy energy. For Southern Ocean relevant parameters the characteristic oscillatory time scale is decadal, with only a weak dependence on the parameter values. The introduction of stochastic wind forcing perturbs the system from equilibrium and leads to relatively large excursions in eddy energy.

There are a number of possible generalisations for the dynamical system. While we have assumed a fixed linear vertical stratification, a fixed but more general vertical stratification can be considered -- see \ref{app:vert}. Since the equations are arrived at by imposing a specific one-dimensional basis for density perturbations, the system is naturally extended by increasing the size of this basis, for example to permit time varying vertical stratification. The channel may then also be coupled to an inter-hemispheric basin, to yield a version of the \citeA{allison2011} model. We note the recent work of \citeA{kobras2022} and \citeA{kobras2024} in the atmospheric context, which derive reduced order models for the two-level quasigeostrophic Phillips model on a $\beta$-plane, and obtain a six or eight dimensional dynamical system respectively. Although our model is simpler, complex models can display a richer structure, such as bifurcation of steady states \cite{kobras2022}, and dependence on eddy geometry \cite{kobras2022,kobras2024}. While an interesting avenue to pursue, an exploration into higher order models is beyond the scope of the present work, which was focused on minimal models for understanding eddy saturation and frictional control.

The idealised model considered here is also simplified by assuming that both the Eulerian and eddy-induced overturnings act on a thin layer at the domain boundaries. That is, both the Eulerian overturning stream function and eddy induced stream function are constant on the interior, and fall rapidly to zero near the boundaries. The steady-state residual circulation is hence zero, and so this model cannot capture eddy compensation effects.

We note that the idealized model presented in this paper provides a dynamical explanation for the ``eddy memory'' mechanism explored in a series of papers by
\citeA{Manucharyan:2017}, \citeA{Moon:2021}, \citeA{Dijkstra:2022} and \citeA{Vanderborght:2024}. It would be of interest to explore the relation between these two approaches and, in particular, the relation and consistency of the oscillatory time scales predicted. 

The discussion of the oscillatory time scale is predicated on the assumption that the deviations from the Hamiltonian case are sufficiently weak. However an increased eddy energy background term, or increased influence of non-linear eddy energy damping, shortens the decay time scale. For the Southern Ocean relevant parameters in Table~\ref{tab:parameters}, with linear energy damping, the decay time scale equals the oscillatory time scale for $\mathcal{E}_0 / \mathcal{E}_* = 9.4$\%, and the oscillatory behaviour is lost entirely from the linearised dynamics at $\mathcal{E}_0 / \mathcal{E}_* \ge 60$\%. In the stochastic case the numerical calculations made use of a weak eddy energy background combined with a strong stochastic wind forcing. Strengthening the eddy energy background term, or weakening the stochastic forcing, each weaken departures from the steady state. Combined, these effects may make observing an oscillatory time scale in a non-idealised setting challenging. However the simple model suggests that, with Southern Ocean relevant parameters, while variations in mean thermal wind transport may be more modest, the eddy energetics may be more sensitive to non-steady wind forcing, due to the potential for rapid growth in eddy energy.

\appendix

\section{Generalised vertical stratification}\label{app:vert}

The density profile in \eqref{eqn:density_1} is generalised to (again defined up to a constant)
\begin{linenomath*}\begin{equation}
  \rho \left( y, z, t \right) = -\frac{\rho_0}{g} m \left( t \right) y - \frac{\rho_0}{g} N_0^2 R \left( z \right),
\end{equation}\end{linenomath*}
where $R \left( z \right)$ is some function defined such that $R \left( -H \right) = -H$ and $R \left( 0 \right) = 0$. We can no longer assume a $z$-independent Gent--McWilliams coefficient, and so instead use
\begin{linenomath*}\begin{equation}
  \kappa_\text{GM} = \alpha E S \left( z \right) \frac{N_0}{\left| m \right|}
\end{equation}\end{linenomath*}
for non-negative $S \left( z \right)$ with maximum value one on $z \in \left[ -H, 0 \right]$.

Repeating the derivation, assuming spatially constant $\alpha$, the non-dimensional dynamical system \eqref{eqn:system_nd} is arrived at as before, but with a modified non-dimensional mean thermal wind transport scale
\begin{linenomath*}\begin{equation}
  \hat{T} = -\frac{1}{2} \frac{1}{\alpha} L H^2 \frac{N_0}{f_0} \lambda \frac{H}{H_R},
\end{equation}\end{linenomath*}
non-dimensional energy scale
\begin{linenomath*}\begin{equation}
  \hat{\mathcal{E}} = -\frac{1}{\alpha} L_x L H \frac{N_0}{f_0} \tau_0 \frac{H}{H_S},
\end{equation}\end{linenomath*}
and non-dimensional angular frequency
\begin{linenomath*}\begin{equation}
  \tilde{\omega}_0^2 = -12 \alpha \frac{1}{L^2} \frac{N_0}{f_0} \frac{\tau_0}{\rho_0} \frac{1}{\lambda^2} \frac{H_R}{H},
\end{equation}\end{linenomath*}
with
\begin{linenomath*}\begin{subequations}\begin{align}
  H_S & = \int_{-H}^0 S \left( z \right) dz, \\
  H_R & = \int_{-H}^0 S \left( z \right) \left( \frac{d R}{d z} \right)^{-1} dz.
\end{align}\end{subequations}\end{linenomath*}


\section*{Data Availability}

Scripts used to generate the figures in this article can be found at \citeA{two_dimensional_a}.

\acknowledgments

JRM, DPM, and JM were supported by the Natural Environment Research Council [NE/R000999/1]. JM also acknowledges financial support from Hong Kong Research Grants Council via grant number 11308021. KM acknowledges the generous support of a David Richards Scholarship at Wadham College, Oxford.

JRM acknowledges helpful discussions with Kostas Zygalakis and Jacques Vanneste.

This work was supported by the Natural Environment Research Council [NE/R000999/1]. This research was funded in whole, or in part, by the Natural Environment Research Council [NE/R000999/1]. For the purpose of open access, the author has applied a creative commons attribution (CC BY) licence to any author accepted manuscript version arising.

\bibliography{references}

\begin{thebibliography}{}

\bibitem [\protect \citeauthoryear {%
Allison%
, Johnson%
\BCBL {}\ \BBA {} Marshall%
}{%
Allison%
\ \protect \BOthers {.}}{%
{\protect \APACyear {2011}}%
}]{%
allison2011}
\APACinsertmetastar {%
allison2011}%
\begin{APACrefauthors}%
Allison, L\BPBI C.%
, Johnson, H\BPBI L.%
\BCBL {}\ \BBA {} Marshall, D\BPBI P.%
\end{APACrefauthors}%
\unskip\
\newblock
\APACrefYearMonthDay{2011}{}{}.
\newblock
{\BBOQ}\APACrefatitle {Spin-up and adjustment of the {A}ntarctic {C}ircumpolar
  {C}urrent and global pycnocline} {Spin-up and adjustment of the {A}ntarctic
  {C}ircumpolar {C}urrent and global pycnocline}.{\BBCQ}
\newblock
\APACjournalVolNumPages{Journal of Marine Research}{69}{2-3}{167--189}.
\PrintBackRefs{\CurrentBib}

\bibitem [\protect \citeauthoryear {%
Ambaum%
\ \BBA {} Novak%
}{%
Ambaum%
\ \BBA {} Novak%
}{%
{\protect \APACyear {2014}}%
}]{%
ambaum2014}
\APACinsertmetastar {%
ambaum2014}%
\begin{APACrefauthors}%
Ambaum, M\BPBI H\BPBI P.%
\BCBT {}\ \BBA {} Novak, L.%
\end{APACrefauthors}%
\unskip\
\newblock
\APACrefYearMonthDay{2014}{}{}.
\newblock
{\BBOQ}\APACrefatitle {A nonlinear oscillator describing storm track
  variability} {A nonlinear oscillator describing storm track
  variability}.{\BBCQ}
\newblock
\APACjournalVolNumPages{Quarterly Journal of the Royal Meteorological
  Society}{140}{685}{2680--2684}.
\newblock
\begin{APACrefDOI} \doi{10.1002/qj.2352} \end{APACrefDOI}
\PrintBackRefs{\CurrentBib}

\bibitem [\protect \citeauthoryear {%
Bachman%
}{%
Bachman%
}{%
{\protect \APACyear {2019}}%
}]{%
bachman2019}
\APACinsertmetastar {%
bachman2019}%
\begin{APACrefauthors}%
Bachman, S\BPBI D.%
\end{APACrefauthors}%
\unskip\
\newblock
\APACrefYearMonthDay{2019}{}{}.
\newblock
{\BBOQ}\APACrefatitle {{The GM+E closure: A framework for coupling backscatter
  with the Gent and McWilliams parameterization}} {{The GM+E closure: A
  framework for coupling backscatter with the Gent and McWilliams
  parameterization}}.{\BBCQ}
\newblock
\APACjournalVolNumPages{Ocean Modell.}{136}{}{85--106}.
\newblock
\begin{APACrefDOI} \doi{10.1016/j.ocemod.2019.02.006} \end{APACrefDOI}
\PrintBackRefs{\CurrentBib}

\bibitem [\protect \citeauthoryear {%
Bagaeva%
, Daninov%
, Oliver%
\BCBL {}\ \BBA {} Juricke%
}{%
Bagaeva%
\ \protect \BOthers {.}}{%
{\protect \APACyear {2020}}%
}]{%
bagaeva2024}
\APACinsertmetastar {%
bagaeva2024}%
\begin{APACrefauthors}%
Bagaeva, E.%
, Daninov, S.%
, Oliver, M.%
\BCBL {}\ \BBA {} Juricke, S.%
\end{APACrefauthors}%
\unskip\
\newblock
\APACrefYearMonthDay{2020}{}{}.
\newblock
{\BBOQ}\APACrefatitle {{Advancing eddy parameterizations: Dynamic energy
  backscatter and the role of subgrid energy advection and stochastic forcing}}
  {{Advancing eddy parameterizations: Dynamic energy backscatter and the role
  of subgrid energy advection and stochastic forcing}}.{\BBCQ}
\newblock
\APACjournalVolNumPages{J. Adv. Model. Earth Syst.}{16}{4}{e2023MS003972}.
\newblock
\begin{APACrefDOI} \doi{10.1029/2023MS003972} \end{APACrefDOI}
\PrintBackRefs{\CurrentBib}

\bibitem [\protect \citeauthoryear {%
Bishop%
\ \protect \BOthers {.}}{%
Bishop%
\ \protect \BOthers {.}}{%
{\protect \APACyear {2016}}%
}]{%
bishop2016}
\APACinsertmetastar {%
bishop2016}%
\begin{APACrefauthors}%
Bishop, S\BPBI P.%
, Gent, P\BPBI R.%
, Bryan, F\BPBI O.%
, Thompson, A\BPBI F.%
, Long, M\BPBI C.%
\BCBL {}\ \BBA {} Abernathey, R\BPBI P.%
\end{APACrefauthors}%
\unskip\
\newblock
\APACrefYearMonthDay{2016}{}{}.
\newblock
{\BBOQ}\APACrefatitle {{Southern Ocean overturning compensation in an
  eddy-resolving climate simulation}} {{Southern Ocean overturning compensation
  in an eddy-resolving climate simulation}}.{\BBCQ}
\newblock
\APACjournalVolNumPages{J. Phys. Oceanogr.}{46}{}{1575--1592}.
\newblock
\begin{APACrefDOI} \doi{10.1175/JPO-D-15-0177.1} \end{APACrefDOI}
\PrintBackRefs{\CurrentBib}

\bibitem [\protect \citeauthoryear {%
Cessi%
}{%
Cessi%
}{%
{\protect \APACyear {2008}}%
}]{%
cessi2008}
\APACinsertmetastar {%
cessi2008}%
\begin{APACrefauthors}%
Cessi, P.%
\end{APACrefauthors}%
\unskip\
\newblock
\APACrefYearMonthDay{2008}{}{}.
\newblock
{\BBOQ}\APACrefatitle {An energy-constrained parameterization of eddy buoyancy
  flux} {An energy-constrained parameterization of eddy buoyancy flux}.{\BBCQ}
\newblock
\APACjournalVolNumPages{Journal of Physical Oceanography}{38}{8}{1807--1819}.
\newblock
\begin{APACrefDOI} \doi{10.1175/2007JPO3812.1} \end{APACrefDOI}
\PrintBackRefs{\CurrentBib}

\bibitem [\protect \citeauthoryear {%
Dijkstra%
, Manucharyan%
\BCBL {}\ \BBA {} Moon%
}{%
Dijkstra%
\ \protect \BOthers {.}}{%
{\protect \APACyear {2022}}%
}]{%
Dijkstra:2022}
\APACinsertmetastar {%
Dijkstra:2022}%
\begin{APACrefauthors}%
Dijkstra, H\BPBI A.%
, Manucharyan, G.%
\BCBL {}\ \BBA {} Moon, W.%
\end{APACrefauthors}%
\unskip\
\newblock
\APACrefYearMonthDay{2022}{}{}.
\newblock
{\BBOQ}\APACrefatitle {Eddy memory in weakly-nonlinear two-layer
  quasi-geostrophic ocean flows} {Eddy memory in weakly-nonlinear two-layer
  quasi-geostrophic ocean flows}.{\BBCQ}
\newblock
\APACjournalVolNumPages{Eur. Phys. J. Plus}{137}{}{1162}.
\PrintBackRefs{\CurrentBib}

\bibitem [\protect \citeauthoryear {%
Eden%
\ \BBA {} Greatbatch%
}{%
Eden%
\ \BBA {} Greatbatch%
}{%
{\protect \APACyear {2008}}%
}]{%
eden2008}
\APACinsertmetastar {%
eden2008}%
\begin{APACrefauthors}%
Eden, C.%
\BCBT {}\ \BBA {} Greatbatch, R\BPBI J.%
\end{APACrefauthors}%
\unskip\
\newblock
\APACrefYearMonthDay{2008}{}{}.
\newblock
{\BBOQ}\APACrefatitle {Towards a mesoscale eddy closure} {Towards a mesoscale
  eddy closure}.{\BBCQ}
\newblock
\APACjournalVolNumPages{Ocean Modelling}{20}{3}{223--239}.
\newblock
\begin{APACrefDOI} \doi{10.1016/j.ocemod.2007.09.002} \end{APACrefDOI}
\PrintBackRefs{\CurrentBib}

\bibitem [\protect \citeauthoryear {%
Farneti%
\ \protect \BOthers {.}}{%
Farneti%
\ \protect \BOthers {.}}{%
{\protect \APACyear {2015}}%
}]{%
farneti2015}
\APACinsertmetastar {%
farneti2015}%
\begin{APACrefauthors}%
Farneti, R.%
, Downes, S\BPBI M.%
, Griffies, S\BPBI M.%
, Marsland, S\BPBI J.%
, Behrens, E.%
, Bentsen, M.%
\BDBL {}Yeager, S\BPBI G.%
\end{APACrefauthors}%
\unskip\
\newblock
\APACrefYearMonthDay{2015}{}{}.
\newblock
{\BBOQ}\APACrefatitle {{An assessment of Antarctic Circumpolar Current and
  Southern Ocean meridional overturning circulation during 1958-2007 in a suite
  of interannual CORE-II simulations}} {{An assessment of Antarctic Circumpolar
  Current and Southern Ocean meridional overturning circulation during
  1958-2007 in a suite of interannual CORE-II simulations}}.{\BBCQ}
\newblock
\APACjournalVolNumPages{Ocean Modell.}{94}{}{84--120}.
\newblock
\begin{APACrefDOI} \doi{10.1016/j.ocemod.2015.07.009} \end{APACrefDOI}
\PrintBackRefs{\CurrentBib}

\bibitem [\protect \citeauthoryear {%
Federer%
}{%
Federer%
}{%
{\protect \APACyear {2021}}%
}]{%
federer2021}
\APACinsertmetastar {%
federer2021}%
\begin{APACrefauthors}%
Federer, M\BPBI P.%
\end{APACrefauthors}%
\unskip\
\newblock
\APACrefYearMonthDay{2021}{}{}.
\newblock
\APACrefbtitle {Predator-prey behaviour of storm tracks.} {Predator-prey
  behaviour of storm tracks.}
\newblock
\APAChowpublished {Master thesis, ETH Zurich}.
\newblock
\begin{APACrefDOI} \doi{10.3929/ethz-b-000690363} \end{APACrefDOI}
\PrintBackRefs{\CurrentBib}

\bibitem [\protect \citeauthoryear {%
Gent%
\ \BBA {} McWilliams%
}{%
Gent%
\ \BBA {} McWilliams%
}{%
{\protect \APACyear {1990}}%
}]{%
gent1990}
\APACinsertmetastar {%
gent1990}%
\begin{APACrefauthors}%
Gent, P\BPBI R.%
\BCBT {}\ \BBA {} McWilliams, J\BPBI C.%
\end{APACrefauthors}%
\unskip\
\newblock
\APACrefYearMonthDay{1990}{}{}.
\newblock
{\BBOQ}\APACrefatitle {Isopycnal mixing in ocean circulation models} {Isopycnal
  mixing in ocean circulation models}.{\BBCQ}
\newblock
\APACjournalVolNumPages{Journal of Physical Oceanography}{20}{1}{150--155}.
\newblock
\begin{APACrefDOI} \doi{10.1175/1520-0485(1990)020<0150:IMIOCM>2.0.CO;2}
  \end{APACrefDOI}
\PrintBackRefs{\CurrentBib}

\bibitem [\protect \citeauthoryear {%
Gent%
, Willebrand%
, McDougall%
\BCBL {}\ \BBA {} McWilliams%
}{%
Gent%
\ \protect \BOthers {.}}{%
{\protect \APACyear {1995}}%
}]{%
gent1995}
\APACinsertmetastar {%
gent1995}%
\begin{APACrefauthors}%
Gent, P\BPBI R.%
, Willebrand, J.%
, McDougall, T\BPBI J.%
\BCBL {}\ \BBA {} McWilliams, J\BPBI C.%
\end{APACrefauthors}%
\unskip\
\newblock
\APACrefYearMonthDay{1995}{}{}.
\newblock
{\BBOQ}\APACrefatitle {Parameterizing eddy-induced tracer transports in ocean
  circulation models} {Parameterizing eddy-induced tracer transports in ocean
  circulation models}.{\BBCQ}
\newblock
\APACjournalVolNumPages{Journal of Physical Oceanography}{25}{4}{463--474}.
\newblock
\begin{APACrefDOI} \doi{10.1175/1520-0485(1995)025<0463:PEITTI>2.0.CO;2}
  \end{APACrefDOI}
\PrintBackRefs{\CurrentBib}

\bibitem [\protect \citeauthoryear {%
Gnanadesikan%
}{%
Gnanadesikan%
}{%
{\protect \APACyear {1999}}%
}]{%
gnanadesikan1999}
\APACinsertmetastar {%
gnanadesikan1999}%
\begin{APACrefauthors}%
Gnanadesikan, A.%
\end{APACrefauthors}%
\unskip\
\newblock
\APACrefYearMonthDay{1999}{}{}.
\newblock
{\BBOQ}\APACrefatitle {A simple predictive model for the structure of the
  oceanic pycnocline} {A simple predictive model for the structure of the
  oceanic pycnocline}.{\BBCQ}
\newblock
\APACjournalVolNumPages{Science}{283}{5410}{2077--2079}.
\newblock
\begin{APACrefDOI} \doi{10.1126/science.283.5410.2077} \end{APACrefDOI}
\PrintBackRefs{\CurrentBib}

\bibitem [\protect \citeauthoryear {%
Hewitt%
\ \protect \BOthers {.}}{%
Hewitt%
\ \protect \BOthers {.}}{%
{\protect \APACyear {2020}}%
}]{%
hewitt2020}
\APACinsertmetastar {%
hewitt2020}%
\begin{APACrefauthors}%
Hewitt, H\BPBI T.%
, Roberts, M.%
, Mathiot, P.%
, Biastoch, A.%
, Blockley, E.%
, Chassignet, E\BPBI P.%
\BDBL {}Zhang, Q.%
\end{APACrefauthors}%
\unskip\
\newblock
\APACrefYearMonthDay{2020}{}{}.
\newblock
{\BBOQ}\APACrefatitle {{Resolving and parameterising the ocean mesoscale in
  earth system models}} {{Resolving and parameterising the ocean mesoscale in
  earth system models}}.{\BBCQ}
\newblock
\APACjournalVolNumPages{Curr. Clim. Change Rep.}{6}{}{137--152}.
\newblock
\begin{APACrefDOI} \doi{10.1007/s40641-020-00164-w} \end{APACrefDOI}
\PrintBackRefs{\CurrentBib}

\bibitem [\protect \citeauthoryear {%
Hughes%
, Meredith%
\BCBL {}\ \BBA {} Heywood%
}{%
Hughes%
\ \protect \BOthers {.}}{%
{\protect \APACyear {1999}}%
}]{%
hughes1999}
\APACinsertmetastar {%
hughes1999}%
\begin{APACrefauthors}%
Hughes, C\BPBI W.%
, Meredith, M\BPBI P.%
\BCBL {}\ \BBA {} Heywood, K\BPBI J.%
\end{APACrefauthors}%
\unskip\
\newblock
\APACrefYearMonthDay{1999}{}{}.
\newblock
{\BBOQ}\APACrefatitle {{Wind-driven transport fluctuations through Drake
  Passage: A southern mode}} {{Wind-driven transport fluctuations through Drake
  Passage: A southern mode}}.{\BBCQ}
\newblock
\APACjournalVolNumPages{J. Phys. Oceanogr.}{29}{}{1971--1992}.
\newblock
\begin{APACrefDOI} \doi{10.1175/1520-0485(1999)029<1971:WDTFTD>2.0.CO;2}
  \end{APACrefDOI}
\PrintBackRefs{\CurrentBib}

\bibitem [\protect \citeauthoryear {%
Jansen%
, Adcroft%
, Khani%
\BCBL {}\ \BBA {} Kong%
}{%
Jansen%
\ \protect \BOthers {.}}{%
{\protect \APACyear {2019}}%
}]{%
jansen2019}
\APACinsertmetastar {%
jansen2019}%
\begin{APACrefauthors}%
Jansen, M\BPBI F.%
, Adcroft, A.%
, Khani, S.%
\BCBL {}\ \BBA {} Kong, H.%
\end{APACrefauthors}%
\unskip\
\newblock
\APACrefYearMonthDay{2019}{}{}.
\newblock
{\BBOQ}\APACrefatitle {{Toward an energetically consistent, resolution aware
  parameterization of ocean mesoscale eddies}} {{Toward an energetically
  consistent, resolution aware parameterization of ocean mesoscale
  eddies}}.{\BBCQ}
\newblock
\APACjournalVolNumPages{J. Adv. Model. Earth Syst.}{1}{}{1--17}.
\newblock
\begin{APACrefDOI} \doi{10.1029/2019MS001750} \end{APACrefDOI}
\PrintBackRefs{\CurrentBib}

\bibitem [\protect \citeauthoryear {%
Jansen%
\ \BBA {} Held%
}{%
Jansen%
\ \BBA {} Held%
}{%
{\protect \APACyear {2014}}%
}]{%
jansen2014}
\APACinsertmetastar {%
jansen2014}%
\begin{APACrefauthors}%
Jansen, M\BPBI F.%
\BCBT {}\ \BBA {} Held, I\BPBI M.%
\end{APACrefauthors}%
\unskip\
\newblock
\APACrefYearMonthDay{2014}{}{}.
\newblock
{\BBOQ}\APACrefatitle {{Parameterizing subgrid-scale eddy effects using
  energetically consistent backscatter}} {{Parameterizing subgrid-scale eddy
  effects using energetically consistent backscatter}}.{\BBCQ}
\newblock
\APACjournalVolNumPages{Ocean Modell.}{80}{}{36--48}.
\newblock
\begin{APACrefDOI} \doi{10.1016/j.ocemod.2014.06.002} \end{APACrefDOI}
\PrintBackRefs{\CurrentBib}

\bibitem [\protect \citeauthoryear {%
Kobras%
, Ambaum%
\BCBL {}\ \BBA {} Lucarini%
}{%
Kobras%
\ \protect \BOthers {.}}{%
{\protect \APACyear {2022}}%
}]{%
kobras2022}
\APACinsertmetastar {%
kobras2022}%
\begin{APACrefauthors}%
Kobras, M.%
, Ambaum, M\BPBI H\BPBI P.%
\BCBL {}\ \BBA {} Lucarini, V.%
\end{APACrefauthors}%
\unskip\
\newblock
\APACrefYearMonthDay{2022}{}{}.
\newblock
{\BBOQ}\APACrefatitle {Eddy saturation in a reduced two-level model of the
  atmosphere} {Eddy saturation in a reduced two-level model of the
  atmosphere}.{\BBCQ}
\newblock
\APACjournalVolNumPages{Geophysical \& Astrophysical Fluid
  Dynamics}{116}{1}{38--55}.
\newblock
\begin{APACrefDOI} \doi{10.1080/03091929.2021.1990912} \end{APACrefDOI}
\PrintBackRefs{\CurrentBib}

\bibitem [\protect \citeauthoryear {%
Kobras%
, Lucarini%
\BCBL {}\ \BBA {} Ambaum%
}{%
Kobras%
\ \protect \BOthers {.}}{%
{\protect \APACyear {2024}}%
}]{%
kobras2024}
\APACinsertmetastar {%
kobras2024}%
\begin{APACrefauthors}%
Kobras, M.%
, Lucarini, V.%
\BCBL {}\ \BBA {} Ambaum, M\BPBI H\BPBI P.%
\end{APACrefauthors}%
\unskip\
\newblock
\APACrefYearMonthDay{2024}{}{}.
\newblock
{\BBOQ}\APACrefatitle {Latitudinal storm track shift in a reduced two-level
  model of the atmosphere} {Latitudinal storm track shift in a reduced
  two-level model of the atmosphere}.{\BBCQ}
\newblock
\APACjournalVolNumPages{Physica D: Nonlinear Phenomena}{458}{}{133926}.
\newblock
\begin{APACrefDOI} \doi{10.1016/j.physd.2023.133926} \end{APACrefDOI}
\PrintBackRefs{\CurrentBib}

\bibitem [\protect \citeauthoryear {%
Lorenz%
}{%
Lorenz%
}{%
{\protect \APACyear {1984}}%
}]{%
lorenz1984}
\APACinsertmetastar {%
lorenz1984}%
\begin{APACrefauthors}%
Lorenz, E\BPBI N.%
\end{APACrefauthors}%
\unskip\
\newblock
\APACrefYearMonthDay{1984}{}{}.
\newblock
{\BBOQ}\APACrefatitle {Irregularity: a fundamental property of the atmosphere}
  {Irregularity: a fundamental property of the atmosphere}.{\BBCQ}
\newblock
\APACjournalVolNumPages{Tellus A}{36A}{2}{98--110}.
\newblock
\begin{APACrefDOI} \doi{10.1111/j.1600-0870.1984.tb00230.x} \end{APACrefDOI}
\PrintBackRefs{\CurrentBib}

\bibitem [\protect \citeauthoryear {%
Maddison%
}{%
Maddison%
}{%
{\protect \APACyear {2024}}%
}]{%
two_dimensional_a}
\APACinsertmetastar {%
two_dimensional_a}%
\begin{APACrefauthors}%
Maddison, J\BPBI R.%
\end{APACrefauthors}%
\unskip\
\newblock
\APACrefYearMonthDay{2024}{}{}.
\newblock
\APACrefbtitle {Scripts for `{A} two-dimensional model for eddy saturation and
  frictional control in the {S}outhern {O}cean' [software].} {Scripts for `{A}
  two-dimensional model for eddy saturation and frictional control in the
  {S}outhern {O}cean' [software].}
\newblock
\APAChowpublished {Zenodo}.
\newblock
\begin{APACrefDOI} \doi{10.5281/zenodo.13365111} \end{APACrefDOI}
\PrintBackRefs{\CurrentBib}

\bibitem [\protect \citeauthoryear {%
Mak%
, Avdis%
\BCBL {}\ \protect \BOthers {.}}{%
Mak%
, Avdis%
\BCBL {}\ \protect \BOthers {.}}{%
{\protect \APACyear {2022}}%
}]{%
mak2022b}
\APACinsertmetastar {%
mak2022b}%
\begin{APACrefauthors}%
Mak, J.%
, Avdis, A.%
, David, T\BPBI W.%
, Lee, H\BPBI S.%
, Na, Y.%
\BCBL {}\ \BBA {} Yan, F\BPBI E.%
\end{APACrefauthors}%
\unskip\
\newblock
\APACrefYearMonthDay{2022}{}{}.
\newblock
{\BBOQ}\APACrefatitle {{On constraining the mesoscale eddy energy dissipation
  time-scale}} {{On constraining the mesoscale eddy energy dissipation
  time-scale}}.{\BBCQ}
\newblock
\APACjournalVolNumPages{J. Adv. Model. Earth Syst.}{14}{}{e2022MS003223}.
\newblock
\begin{APACrefDOI} \doi{10.1029/2022MS003223} \end{APACrefDOI}
\PrintBackRefs{\CurrentBib}

\bibitem [\protect \citeauthoryear {%
Mak%
, Maddison%
, Marshall%
\BCBL {}\ \BBA {} Munday%
}{%
Mak%
\ \protect \BOthers {.}}{%
{\protect \APACyear {2018}}%
}]{%
mak2018}
\APACinsertmetastar {%
mak2018}%
\begin{APACrefauthors}%
Mak, J.%
, Maddison, J\BPBI R.%
, Marshall, D\BPBI P.%
\BCBL {}\ \BBA {} Munday, D\BPBI R.%
\end{APACrefauthors}%
\unskip\
\newblock
\APACrefYearMonthDay{2018}{}{}.
\newblock
{\BBOQ}\APACrefatitle {Implementation of a geometrically informed and
  energetically constrained mesoscale eddy parameterization in an ocean
  circulation model} {Implementation of a geometrically informed and
  energetically constrained mesoscale eddy parameterization in an ocean
  circulation model}.{\BBCQ}
\newblock
\APACjournalVolNumPages{Journal of Physical Oceanography}{48}{10}{2363--2382}.
\newblock
\begin{APACrefDOI} \doi{10.1175/JPO-D-18-0017.1} \end{APACrefDOI}
\PrintBackRefs{\CurrentBib}

\bibitem [\protect \citeauthoryear {%
Mak%
, Marshall%
, Maddison%
\BCBL {}\ \BBA {} Bachman%
}{%
Mak%
\ \protect \BOthers {.}}{%
{\protect \APACyear {2017}}%
}]{%
mak2017}
\APACinsertmetastar {%
mak2017}%
\begin{APACrefauthors}%
Mak, J.%
, Marshall, D\BPBI P.%
, Maddison, J\BPBI R.%
\BCBL {}\ \BBA {} Bachman, S\BPBI D.%
\end{APACrefauthors}%
\unskip\
\newblock
\APACrefYearMonthDay{2017}{}{}.
\newblock
{\BBOQ}\APACrefatitle {Emergent eddy saturation from an energy constrained eddy
  parameterisation} {Emergent eddy saturation from an energy constrained eddy
  parameterisation}.{\BBCQ}
\newblock
\APACjournalVolNumPages{Ocean Modelling}{112}{}{125--138}.
\newblock
\begin{APACrefDOI} \doi{10.1016/j.ocemod.2017.02.007} \end{APACrefDOI}
\PrintBackRefs{\CurrentBib}

\bibitem [\protect \citeauthoryear {%
Mak%
, Marshall%
, Madec%
\BCBL {}\ \BBA {} Maddison%
}{%
Mak%
, Marshall%
\BCBL {}\ \protect \BOthers {.}}{%
{\protect \APACyear {2022}}%
}]{%
mak2022}
\APACinsertmetastar {%
mak2022}%
\begin{APACrefauthors}%
Mak, J.%
, Marshall, D\BPBI P.%
, Madec, G.%
\BCBL {}\ \BBA {} Maddison, J\BPBI R.%
\end{APACrefauthors}%
\unskip\
\newblock
\APACrefYearMonthDay{2022}{}{}.
\newblock
{\BBOQ}\APACrefatitle {{Acute sensitivity of global ocean circulation and heat
  content to eddy energy dissipation time-scale}} {{Acute sensitivity of global
  ocean circulation and heat content to eddy energy dissipation
  time-scale}}.{\BBCQ}
\newblock
\APACjournalVolNumPages{Geophys. Res. Lett.}{49}{}{e2021GL097259}.
\newblock
\begin{APACrefDOI} \doi{10.1029/2021GL097259} \end{APACrefDOI}
\PrintBackRefs{\CurrentBib}

\bibitem [\protect \citeauthoryear {%
Manucharyan%
, Thompson%
\BCBL {}\ \BBA {} Spall%
}{%
Manucharyan%
\ \protect \BOthers {.}}{%
{\protect \APACyear {2017}}%
}]{%
Manucharyan:2017}
\APACinsertmetastar {%
Manucharyan:2017}%
\begin{APACrefauthors}%
Manucharyan, G\BPBI E.%
, Thompson, A\BPBI F.%
\BCBL {}\ \BBA {} Spall, M\BPBI A.%
\end{APACrefauthors}%
\unskip\
\newblock
\APACrefYearMonthDay{2017}{}{}.
\newblock
{\BBOQ}\APACrefatitle {Eddy memory mode of multidecadal variability in
  residual-mean ocean circulations with application to the Beaufort Gyre} {Eddy
  memory mode of multidecadal variability in residual-mean ocean circulations
  with application to the beaufort gyre}.{\BBCQ}
\newblock
\APACjournalVolNumPages{J. Phys. Oceanogr.}{47}{}{855--866}.
\PrintBackRefs{\CurrentBib}

\bibitem [\protect \citeauthoryear {%
Marshall%
\ \BBA {} Adcroft%
}{%
Marshall%
\ \BBA {} Adcroft%
}{%
{\protect \APACyear {2010}}%
}]{%
marshall2010}
\APACinsertmetastar {%
marshall2010}%
\begin{APACrefauthors}%
Marshall, D\BPBI P.%
\BCBT {}\ \BBA {} Adcroft, A\BPBI J.%
\end{APACrefauthors}%
\unskip\
\newblock
\APACrefYearMonthDay{2010}{}{}.
\newblock
{\BBOQ}\APACrefatitle {Parameterization of ocean eddies: Potential vorticity
  mixing, energetics and {A}rnold's first stability theorem} {Parameterization
  of ocean eddies: Potential vorticity mixing, energetics and {A}rnold's first
  stability theorem}.{\BBCQ}
\newblock
\APACjournalVolNumPages{Ocean Modelling}{32}{3}{188--204}.
\newblock
\begin{APACrefDOI} \doi{10.1016/j.ocemod.2010.02.001} \end{APACrefDOI}
\PrintBackRefs{\CurrentBib}

\bibitem [\protect \citeauthoryear {%
Marshall%
, Ambaum%
, Maddison%
, Munday%
\BCBL {}\ \BBA {} Novak%
}{%
Marshall%
\ \protect \BOthers {.}}{%
{\protect \APACyear {2017}}%
}]{%
marshall2017}
\APACinsertmetastar {%
marshall2017}%
\begin{APACrefauthors}%
Marshall, D\BPBI P.%
, Ambaum, M\BPBI H\BPBI P.%
, Maddison, J\BPBI R.%
, Munday, D\BPBI R.%
\BCBL {}\ \BBA {} Novak, L.%
\end{APACrefauthors}%
\unskip\
\newblock
\APACrefYearMonthDay{2017}{}{}.
\newblock
{\BBOQ}\APACrefatitle {Eddy saturation and frictional control of the
  {A}ntarctic {C}ircumpolar {C}urrent} {Eddy saturation and frictional control
  of the {A}ntarctic {C}ircumpolar {C}urrent}.{\BBCQ}
\newblock
\APACjournalVolNumPages{Geophysical Research Letters}{44}{1}{286--292}.
\newblock
\begin{APACrefDOI} \doi{10.1002/2016GL071702} \end{APACrefDOI}
\PrintBackRefs{\CurrentBib}

\bibitem [\protect \citeauthoryear {%
Marshall%
, Maddison%
\BCBL {}\ \BBA {} Berloff%
}{%
Marshall%
\ \protect \BOthers {.}}{%
{\protect \APACyear {2012}}%
}]{%
marshall2012}
\APACinsertmetastar {%
marshall2012}%
\begin{APACrefauthors}%
Marshall, D\BPBI P.%
, Maddison, J\BPBI R.%
\BCBL {}\ \BBA {} Berloff, P\BPBI S.%
\end{APACrefauthors}%
\unskip\
\newblock
\APACrefYearMonthDay{2012}{}{}.
\newblock
{\BBOQ}\APACrefatitle {A framework for parameterizing eddy potential vorticity
  fluxes} {A framework for parameterizing eddy potential vorticity
  fluxes}.{\BBCQ}
\newblock
\APACjournalVolNumPages{Journal of Physical Oceanography}{42}{4}{539--557}.
\newblock
\begin{APACrefDOI} \doi{10.1175/JPO-D-11-048.1} \end{APACrefDOI}
\PrintBackRefs{\CurrentBib}

\bibitem [\protect \citeauthoryear {%
Moon%
, Manucharyan%
\BCBL {}\ \BBA {} Dijkstra%
}{%
Moon%
\ \protect \BOthers {.}}{%
{\protect \APACyear {2021}}%
}]{%
Moon:2021}
\APACinsertmetastar {%
Moon:2021}%
\begin{APACrefauthors}%
Moon, W.%
, Manucharyan, G\BPBI E.%
\BCBL {}\ \BBA {} Dijkstra, H\BPBI A.%
\end{APACrefauthors}%
\unskip\
\newblock
\APACrefYearMonthDay{2021}{}{}.
\newblock
{\BBOQ}\APACrefatitle {Eddy memory as an explanation of intraseasonal periodic
  behaviour in baroclinic eddies} {Eddy memory as an explanation of
  intraseasonal periodic behaviour in baroclinic eddies}.{\BBCQ}
\newblock
\APACjournalVolNumPages{Quart. J. Roy. Meteor. Soc.}{147}{}{2395--2408}.
\PrintBackRefs{\CurrentBib}

\bibitem [\protect \citeauthoryear {%
Munday%
, Johnson%
\BCBL {}\ \BBA {} Marshall%
}{%
Munday%
\ \protect \BOthers {.}}{%
{\protect \APACyear {2013}}%
}]{%
munday2013}
\APACinsertmetastar {%
munday2013}%
\begin{APACrefauthors}%
Munday, D\BPBI R.%
, Johnson, H\BPBI L.%
\BCBL {}\ \BBA {} Marshall, D\BPBI P.%
\end{APACrefauthors}%
\unskip\
\newblock
\APACrefYearMonthDay{2013}{}{}.
\newblock
{\BBOQ}\APACrefatitle {Eddy saturation of equilibrated circumpolar currents}
  {Eddy saturation of equilibrated circumpolar currents}.{\BBCQ}
\newblock
\APACjournalVolNumPages{Journal of Physical Oceanography}{43}{3}{507--532}.
\newblock
\begin{APACrefDOI} \doi{10.1175/JPO-D-12-095.1} \end{APACrefDOI}
\PrintBackRefs{\CurrentBib}

\bibitem [\protect \citeauthoryear {%
Novak%
, Ambaum%
\BCBL {}\ \BBA {} Harvey%
}{%
Novak%
\ \protect \BOthers {.}}{%
{\protect \APACyear {2018}}%
}]{%
novak2018}
\APACinsertmetastar {%
novak2018}%
\begin{APACrefauthors}%
Novak, L.%
, Ambaum, M\BPBI H\BPBI P.%
\BCBL {}\ \BBA {} Harvey, B\BPBI J.%
\end{APACrefauthors}%
\unskip\
\newblock
\APACrefYearMonthDay{2018}{}{}.
\newblock
{\BBOQ}\APACrefatitle {Baroclinic adjustment and dissipative control of storm
  tracks} {Baroclinic adjustment and dissipative control of storm
  tracks}.{\BBCQ}
\newblock
\APACjournalVolNumPages{Journal of the Atmospheric
  Sciences}{75}{9}{2955--2970}.
\newblock
\begin{APACrefDOI} \doi{10.1175/JAS-D-17-0210.1} \end{APACrefDOI}
\PrintBackRefs{\CurrentBib}

\bibitem [\protect \citeauthoryear {%
Novak%
, Ambaum%
\BCBL {}\ \BBA {} Tailleux%
}{%
Novak%
\ \protect \BOthers {.}}{%
{\protect \APACyear {2017}}%
}]{%
novak2017}
\APACinsertmetastar {%
novak2017}%
\begin{APACrefauthors}%
Novak, L.%
, Ambaum, M\BPBI H\BPBI P.%
\BCBL {}\ \BBA {} Tailleux, R.%
\end{APACrefauthors}%
\unskip\
\newblock
\APACrefYearMonthDay{2017}{}{}.
\newblock
{\BBOQ}\APACrefatitle {Marginal stability and predator--prey behaviour within
  storm tracks} {Marginal stability and predator--prey behaviour within storm
  tracks}.{\BBCQ}
\newblock
\APACjournalVolNumPages{Quarterly Journal of the Royal Meteorological
  Society}{143}{704}{1421--1433}.
\newblock
\begin{APACrefDOI} \doi{10.1002/qj.3014} \end{APACrefDOI}
\PrintBackRefs{\CurrentBib}

\bibitem [\protect \citeauthoryear {%
Olbers%
\ \BBA {} Lettmann%
}{%
Olbers%
\ \BBA {} Lettmann%
}{%
{\protect \APACyear {2007}}%
}]{%
olbers2007}
\APACinsertmetastar {%
olbers2007}%
\begin{APACrefauthors}%
Olbers, D.%
\BCBT {}\ \BBA {} Lettmann, K.%
\end{APACrefauthors}%
\unskip\
\newblock
\APACrefYearMonthDay{2007}{}{}.
\newblock
{\BBOQ}\APACrefatitle {{Barotropic and baroclinic processes in the transport
  variability of the Antarctic Circumpolar Current}} {{Barotropic and
  baroclinic processes in the transport variability of the Antarctic
  Circumpolar Current}}.{\BBCQ}
\newblock
\APACjournalVolNumPages{Ocean Dyn.}{57}{}{559--578}.
\newblock
\begin{APACrefDOI} \doi{10.1007/s10236-007-0126-1} \end{APACrefDOI}
\PrintBackRefs{\CurrentBib}

\bibitem [\protect \citeauthoryear {%
Ong%
, Doddridge%
, Constantinou%
, Hogg%
\BCBL {}\ \BBA {} England%
}{%
Ong%
\ \protect \BOthers {.}}{%
{\protect \APACyear {2024}}%
}]{%
ong2024}
\APACinsertmetastar {%
ong2024}%
\begin{APACrefauthors}%
Ong, E\BPBI Q\BPBI Y.%
, Doddridge, E.%
, Constantinou, N\BPBI C.%
, Hogg, A\BPBI M.%
\BCBL {}\ \BBA {} England, M\BPBI H.%
\end{APACrefauthors}%
\unskip\
\newblock
\APACrefYearMonthDay{2024}{}{}.
\newblock
{\BBOQ}\APACrefatitle {Intrinsically episodic {A}ntarctic shelf intrusions of
  {C}ircumpolar {D}eep {W}ater via canyons} {Intrinsically episodic {A}ntarctic
  shelf intrusions of {C}ircumpolar {D}eep {W}ater via canyons}.{\BBCQ}
\newblock
\APACjournalVolNumPages{Journal of Physical Oceanography}{54}{5}{1195--1210}.
\newblock
\begin{APACrefDOI} \doi{10.1175/JPO-D-23-0067.1} \end{APACrefDOI}
\PrintBackRefs{\CurrentBib}

\bibitem [\protect \citeauthoryear {%
Pavliotis%
}{%
Pavliotis%
}{%
{\protect \APACyear {2014}}%
}]{%
pavliotis2014}
\APACinsertmetastar {%
pavliotis2014}%
\begin{APACrefauthors}%
Pavliotis, G\BPBI A.%
\end{APACrefauthors}%
\unskip\
\newblock
\APACrefYear{2014}.
\newblock
\APACrefbtitle {Stochastic processes and applications} {Stochastic processes
  and applications}.
\newblock
\APACaddressPublisher{}{Springer Science+Business Media New York}.
\PrintBackRefs{\CurrentBib}

\bibitem [\protect \citeauthoryear {%
Pedlosky%
}{%
Pedlosky%
}{%
{\protect \APACyear {1970}}%
}]{%
pedlosky1970}
\APACinsertmetastar {%
pedlosky1970}%
\begin{APACrefauthors}%
Pedlosky, J.%
\end{APACrefauthors}%
\unskip\
\newblock
\APACrefYearMonthDay{1970}{}{}.
\newblock
{\BBOQ}\APACrefatitle {Finite-amplitude baroclinic waves} {Finite-amplitude
  baroclinic waves}.{\BBCQ}
\newblock
\APACjournalVolNumPages{Journal of the Atmospheric Sciences}{27}{1}{15--30}.
\newblock
\begin{APACrefDOI} \doi{10.1175/1520-0469(1970)027<0015:FABW>2.0.CO;2}
  \end{APACrefDOI}
\PrintBackRefs{\CurrentBib}

\bibitem [\protect \citeauthoryear {%
Sinha%
\ \BBA {} Abernathey%
}{%
Sinha%
\ \BBA {} Abernathey%
}{%
{\protect \APACyear {2016}}%
}]{%
sinha2016}
\APACinsertmetastar {%
sinha2016}%
\begin{APACrefauthors}%
Sinha, A.%
\BCBT {}\ \BBA {} Abernathey, R\BPBI P.%
\end{APACrefauthors}%
\unskip\
\newblock
\APACrefYearMonthDay{2016}{}{}.
\newblock
{\BBOQ}\APACrefatitle {{Time scales of Southern Ocean eddy equilibration}}
  {{Time scales of Southern Ocean eddy equilibration}}.{\BBCQ}
\newblock
\APACjournalVolNumPages{J. Phys. Oceanogr.}{46}{}{2785--2805}.
\newblock
\begin{APACrefDOI} \doi{10.1175/JPO-D-16-0041.1} \end{APACrefDOI}
\PrintBackRefs{\CurrentBib}

\bibitem [\protect \citeauthoryear {%
Straub%
}{%
Straub%
}{%
{\protect \APACyear {1993}}%
}]{%
straub1993}
\APACinsertmetastar {%
straub1993}%
\begin{APACrefauthors}%
Straub, D\BPBI N.%
\end{APACrefauthors}%
\unskip\
\newblock
\APACrefYearMonthDay{1993}{}{}.
\newblock
{\BBOQ}\APACrefatitle {On the transport and angular momentum balance of channel
  models of the {A}ntarctic {C}ircumpolar {C}urrent} {On the transport and
  angular momentum balance of channel models of the {A}ntarctic {C}ircumpolar
  {C}urrent}.{\BBCQ}
\newblock
\APACjournalVolNumPages{Journal of Physical Oceanography}{23}{4}{776--782}.
\newblock
\begin{APACrefDOI} \doi{10.1175/1520-0485(1993)023<0776:OTTAAM>2.0.CO;2}
  \end{APACrefDOI}
\PrintBackRefs{\CurrentBib}

\bibitem [\protect \citeauthoryear {%
Sura%
\ \BBA {} Gille%
}{%
Sura%
\ \BBA {} Gille%
}{%
{\protect \APACyear {2003}}%
}]{%
sura2003}
\APACinsertmetastar {%
sura2003}%
\begin{APACrefauthors}%
Sura, P.%
\BCBT {}\ \BBA {} Gille, S\BPBI T.%
\end{APACrefauthors}%
\unskip\
\newblock
\APACrefYearMonthDay{2003}{}{}.
\newblock
{\BBOQ}\APACrefatitle {{Interpreting wind-driven Southern Ocean variability in
  a stochastic framework}} {{Interpreting wind-driven Southern Ocean
  variability in a stochastic framework}}.{\BBCQ}
\newblock
\APACjournalVolNumPages{J. Mar. Res.}{61}{}{313--334}.
\newblock
\begin{APACrefDOI} \doi{10.1357/002224003322201214} \end{APACrefDOI}
\PrintBackRefs{\CurrentBib}

\bibitem [\protect \citeauthoryear {%
A\BPBI F.~Thompson%
, Stewart%
\BCBL {}\ \BBA {} Bischoff%
}{%
A\BPBI F.~Thompson%
\ \protect \BOthers {.}}{%
{\protect \APACyear {2016}}%
}]{%
thompson2016}
\APACinsertmetastar {%
thompson2016}%
\begin{APACrefauthors}%
Thompson, A\BPBI F.%
, Stewart, A\BPBI L.%
\BCBL {}\ \BBA {} Bischoff, T.%
\end{APACrefauthors}%
\unskip\
\newblock
\APACrefYearMonthDay{2016}{}{}.
\newblock
{\BBOQ}\APACrefatitle {{A multibasin residual-mean model for the global
  overturning circulation}} {{A multibasin residual-mean model for the global
  overturning circulation}}.{\BBCQ}
\newblock
\APACjournalVolNumPages{J. Phys. Oceanogr.}{46}{}{2583--2604}.
\newblock
\begin{APACrefDOI} \doi{10.1175/JPO-D-15-0204.1} \end{APACrefDOI}
\PrintBackRefs{\CurrentBib}

\bibitem [\protect \citeauthoryear {%
P\BPBI D.~Thompson%
}{%
P\BPBI D.~Thompson%
}{%
{\protect \APACyear {1987}}%
}]{%
thompson1987}
\APACinsertmetastar {%
thompson1987}%
\begin{APACrefauthors}%
Thompson, P\BPBI D.%
\end{APACrefauthors}%
\unskip\
\newblock
\APACrefYearMonthDay{1987}{}{}.
\newblock
{\BBOQ}\APACrefatitle {Large-scale dynamical response to differential heating:
  statistical equilibrium states and amplitude vacillation} {Large-scale
  dynamical response to differential heating: statistical equilibrium states
  and amplitude vacillation}.{\BBCQ}
\newblock
\APACjournalVolNumPages{Journal of the Atmospheric
  Sciences}{44}{8}{1237--1248}.
\newblock
\begin{APACrefDOI} \doi{10.1175/1520-0469(1987)044<1237:LSDRTD>2.0.CO;2}
  \end{APACrefDOI}
\PrintBackRefs{\CurrentBib}

\bibitem [\protect \citeauthoryear {%
Vanderborght%
, Demaeyer%
, Manucharyan%
, Moon%
\BCBL {}\ \BBA {} Dijkstra%
}{%
Vanderborght%
\ \protect \BOthers {.}}{%
{\protect \APACyear {2024}}%
}]{%
Vanderborght:2024}
\APACinsertmetastar {%
Vanderborght:2024}%
\begin{APACrefauthors}%
Vanderborght, E.%
, Demaeyer, J.%
, Manucharyan, G.%
, Moon, W.%
\BCBL {}\ \BBA {} Dijkstra, H\BPBI A.%
\end{APACrefauthors}%
\unskip\
\newblock
\APACrefYearMonthDay{2024}{}{}.
\newblock
{\BBOQ}\APACrefatitle {Physics of the Eddy Memory Kernel of a Baroclinic
  Midlatitude Atmosphere} {Physics of the eddy memory kernel of a baroclinic
  midlatitude atmosphere}.{\BBCQ}
\newblock
\APACjournalVolNumPages{J. Atmos. Sci.}{81}{}{691--711}.
\PrintBackRefs{\CurrentBib}

\bibitem [\protect \citeauthoryear {%
Yankovsky%
, Bachman%
, Smith%
\BCBL {}\ \BBA {} Zanna%
}{%
Yankovsky%
\ \protect \BOthers {.}}{%
{\protect \APACyear {2024}}%
}]{%
yankovsky2024}
\APACinsertmetastar {%
yankovsky2024}%
\begin{APACrefauthors}%
Yankovsky, E.%
, Bachman, S\BPBI D.%
, Smith, K\BPBI S.%
\BCBL {}\ \BBA {} Zanna, L.%
\end{APACrefauthors}%
\unskip\
\newblock
\APACrefYearMonthDay{2024}{}{}.
\newblock
{\BBOQ}\APACrefatitle {{Vertical structure and energetic constraints for a
  backscatter parameterization of ocean mesoscale eddies}} {{Vertical structure
  and energetic constraints for a backscatter parameterization of ocean
  mesoscale eddies}}.{\BBCQ}
\newblock
\APACjournalVolNumPages{J. Adv. Model. Earth. Syst.}{16}{7}{e2023MS004093}.
\newblock
\begin{APACrefDOI} \doi{10.1029/2023MS004093} \end{APACrefDOI}
\PrintBackRefs{\CurrentBib}

\end{thebibliography}

\end{document}